\documentclass[a4paper,11pt]{article}
\pdfoutput=1 

\usepackage{jcappub} 

\usepackage[T1]{fontenc} 
\usepackage{array}
\usepackage{tabularx}
\usepackage{booktabs}
\newcolumntype{Y}{>{\centering\arraybackslash}X}
\usepackage{subcaption}

\title{\boldmath Two-sector leptogenesis in a two-Higgs-doublet model with
spontaneous CP violation}

\author[a]{Debashree Priyadarsini Das,}
\author[b]{ Joy Ganguly}
\author[a,1]{and Sasmita Mishra,\note{Corresponding author.}}

\affiliation[a]{Department of Physics and Astronomy, National Institute of Technology Rourkela, Sundargarh, Odisha, India, $769008$}
\affiliation[b]{Department of BSH, Institute of Engineering and Management Kolkata, University of Engineering and Management Kolkata, Kolkata, India, $700160$}

\emailAdd{debashreepriyadarsini\_das@nitrkl.ac.in}
\emailAdd{mishras@nitrkl.ac.in}
\emailAdd{joyganguli2013@gmail.com}

\abstract{The extension of the Standard Model (SM) field content with one inert Higgs doublet (IHD) and three right-handed neutrinos (RHNs) is a well-motivated approach. The key advantages of the model include the appearance of a weakly interacting massive particle (WIMP) like dark matter (DM) candidate from the neutral component of the IHD, along with the plausible explanation of the sub-eV mass range of SM neutrinos via the radiative seesaw mechanism. Additionally, the decay of RHNs can contextualize the baryon asymmetry of the universe via leptogenesis and is intricately connected to CP violation. Also, given the ongoing searches for light scalars at various experimental facilities, the extended Higgs sector of the model continues to be at the forefront. However, this scotogenic framework encounters a deficiency in providing the observed amount of relic density for a particular mass range $\sim (80 - 500) $ GeV of its DM candidate, hence requiring further augmentation. Also, the WIMP scenarios have not yet resulted in conclusive hints at the direct detection experiments. In this context, our work is based on further extension of the above Scotogenic model by a dark sector. Additionally, considering the cosmic coincidence aspect, we operate within the framework of two-sector leptogenesis. To have a predictive flavor structure in the visible sector, we impose $A_4$ symmetry. Also, we adhere to spontaneous CP violation via complex vacuum expectation value of the falvon field, leading to a situation where there is only one CP-violating phase as a common connection between the visible and dark sectors. In our analysis, we find for the lightest RHN mass $\sim 10^{10}$ GeV, our results are in good agreement with the observational ratio of relic densities, i.e., $\Omega_{\rm DM}/\Omega_{\rm b} \sim 5$ for a few GeV range of mass of the dark sector DM candidate. }


\begin{document}
\maketitle
\flushbottom

\section{Introduction}
 \label{sec:intro}
Despite its enormous success, the Standard Model falls short of being a complete theory, as it can not account for certain observations like neutrino oscillation \cite{Fukuda:1998mi, Ahmad:2002jz, Ahmad:2002ka, Bahcall:2004mz} and the existence of their non-zero, but tiny mass. Their confirmed empirical proofs invoke the extension of the particle content of the SM. In addition to this, a handful of experimental facts from cosmology can not be accommodated within the SM, such as the critical densities of dark matter and baryonic matter, 
$\Omega_{\rm DM} h^2 = 0.120 \pm 0.001$ and 
$\Omega_{\rm b} h^2 = 0.0224 \pm 0.0001$
at $68\%$ CL \cite{Planck:2018vyg}, respectively, where $h$ is the Hubble parameter in units of $100 ~{\rm km}~ s^{-1} {\rm Mpc}^{-1}$. The resulting ratio of densities
$\Omega_{\rm DM}/\Omega_{\rm b} =4.83 \pm 0.87$ is in the same order. Whereas the theoretical explanation of non-zero neutrino mass via the seesaw mechanism predicts the existence of heavy states beyond the SM, the decays of the same heavy states in the early Universe can produce the baryon asymmetry of the Universe via leptogenesis \cite{Fukugita:1986hr}. The apparent link among all the above phenomena fascinates us to investigate the possibility of whether a unified explanation exists by going beyond the SM. In this work, we study the unified explanation of the aforementioned phenomenologies in a variant of the two-Higgs-doublet model ($2$HDM) with a two-sector leptogenesis scenario with spontaneous charge parity (CP) violation.
 
An attractive scenario is when the content of SM matter is complemented by an IHD, denoted as $\eta$, along with three RHNs $N_{R_i}$. In this context, $2$HDM  (Ref.\cite{Branco:2011iw} for a review)   emerges as a promising extension with its simplicity and rich phenomenology, offering the neutral component of the IHD as a stable dark matter candidate, resembling a WIMP candidate. The concept of IHD was initially proposed by Deshpande and Ma \cite{Deshpande:1977rw} to study the electroweak symmetry breaking pattern.  On the other hand, the addition of RHNs can simultaneously provide a plausible explanation for the sub-eV mass range of SM neutrinos via the type-I seesaw mechanism \cite{Mohapatra:1974gc, Minkowski:1977sc, Senjanovic:1975rk, Mohapatra:1979ia, Schechter:1980gr, Gell-Mann:1979vob, Glashow:1979nm, Yanagida:1979as}, and their decays in the early universe can act as a source of baryon asymmetry of the universe (BAU) via leptogenesis. 
The neutral component of IHD is an ideal candidate for WIMP, providing significant observational hints, as studied in references \cite{LopezHonorez:2006gr, Gustafsson:2007pc}. However, in references \cite{Banerjee:2019luv, LopezHonorez:2006gr, Astros:2023gda} it was shown that for the 
mass range of the IHD $\sim(80 - 500)$ GeV, the relic density of DM remains underabundant. For the DM mass greater than $80$ GeV, it annihilates into $Z$ and $W$ boson pairs, and for the mass greater than the SM Higgs mass, the
annihilation channel opens up to a pair of scalars. 
The WIMP scenarios, for the low-mass regime, are highly constrained at direct detection experiments: XENON1T \cite{XENON:2019gfn}, 
LUX-ZEPLIN \cite{LZ:2022lsv,LZCollaboration:2024lux}, and DARWIN \cite{DARWIN:2016hyl} and indirect detection:
FERMI-LAT \cite{Fermi-LAT:2016uux}, AMS \cite{AMS:2002yni}, and H.E.S.S \cite{Abdalla:2016olq}. For the high-mass regime, the direct production of DM at colliders will be suppressed compared to the low-mass regime. So, based on
collider searches and detection experiments, the DM mass regime $\lesssim 500$ GeV appears to be severely constrained.

Among various scalar extensions of the SM, the $2$HDM is well motivated since the structure resembles that of 
other new physics models such as supersymmetric extension of the SM, grand unified theories, axion, and composite Higgs models (for 
a review see Ref. \cite{Branco:2011iw}).
The extended scalar structure of $2$HDM is motivated by various
experimental searches for new scalar bosons.
The SM hosts one scalar doublet, but multiple fermion doublets and fermion mixings. This hints towards looking for
an extended scalar sector. The possible indication comes from the constraints from the $\rho$-parameter at tree level \cite{Langacker:1980js}. If there are $n$ scalars, their weak isospins,
weak hypercharges, and vacuum expectation values (VEVs) 
contribute to the $\rho$-parameter. The experimental value 
$\rho= 1$ is compatible with models with extended scalar sector \cite{Chanowitz:1985ug}. From the results reported by the ATLAS and CMS collaborations 
at the Large Hadron Collider (LHC) on the discovery of a scalar resembling the SM Higgs boson \cite{ATLAS:2012yve, CMS:2012qbp}, there have been no claims of 
any new resonance at $5\sigma$ level. However, a local excess of charged Higgs boson has been reported with masses between $60$ and $160$ GeV \cite{ATLAS:2023bzb} and that between $70$ and $110$ GeV \cite{CMS:2018cyk}.
Some future searches by the Belle collaboration \cite{Belle:2021rcl} and ATLAS collaboration \cite{ATLAS:2023tkl} are in line. Hence, models based on $2$HDM continue to be testing grounds for phenomenological studies.

Whereas leptogenesis and the generation of light neutrino mass are very well explained in the $2$HDM, the DM scenario seems to be under tension. But the scalar spectrum of the theory remains promising 
as per experimental searches.  So one could think of the 
dark sector hosting multiple candidates for DM and a mechanism of DM genesis similar to the visible sector.
In this way, it would be possible to evade direct and 
indirect detection limits. Inspired by the cosmic coincidence 
($\Omega_{\rm DM}/ \Omega_b \sim 5$) the asymmetric dark matter 
(ADM) scenarios offer a promising direction, where an asymmetry is created either in the dark or the 
visible sector, and at later times, the asymmetry is transferred to 
the other sector via contact interactions \cite{Abada:2007ux,Kaplan:2009ag,Cohen:2009fz,Feldstein:2010xe,Haba:2010bm,
Boucenna:2013wba,Frandsen:2018jfi}. In this case,
the number densities of baryons and DM, $n_{\rm DM}, n_{\rm b}$, are in a similar range ($n_{\rm DM}\sim n_{\rm b}$)
resulting in the DM mass in the GeV ballpark. In a bit different approach, called two-sector leptogenesis \cite{Falkowski:2011xh}, where the lepton and
DM asymmetries can be generated simultaneously at high temperature
and can naturally lead to a large hierarchy between $n_{\rm DM}$ and 
$n_{\rm b}$. In this scenario the DM masses can be in keV to TeV. In this paper, we work
in the framework of two-sector leptogenesis by augmenting the visible sector with
a dark sector consisting of one fermion and one scalar.

The CP violation required for the two-sector leptogenesis
 can come from two sources: (i) explicit CP violation originating from the 
 complex Yukawa couplings, (ii) spontaneous CP violation originating from the 
 complex VEV of a scalar field leading to certain symmetry breaking. The latter 
 is more economical in terms of parameter counting and is theoretically well 
 motivated  \cite{Lee:1974jb}. Also, the phase of the complex VEV can act as 
 a common link between low- and high-scale CP violation, such as in neutrino 
 oscillation and two-sector leptogenesis. In this work, we extend the SM 
 particle content and the gauge group with fields charged under a flavor symmetry, and choose the symmetry group to be $A_4$. The extension leads to a predictive
 fermion mixing structure compatible with neutrino oscillation data. 
 Also, we choose one flavon developing a complex VEV, which leads to spontaneous
 CP violation ensures a reduction in the number of free parameters in the model.
 
The paper is organised as follows. In section (\ref{model}), we lay out the model
consisting of the visible and dark sectors. We discuss the CP violation arising from the 
VEV of the flavon fields due to $A_4$ symmetry breaking. We also discuss the origin of
charged lepton and neutrino mass matrices. 
In section (\ref{adm-bau}), we discuss the framework of two-sector leptogenesis. We present the results
and conclusion in sections (\ref{result}) and (\ref{conclusion}) respectively. 
\section{The Model}
\label{model}
The particle content of the model is divided into two
sectors: visible and hidden. In the visible sector, the model
resembles the particle content of the $2{\rm HDM}$
based on the SM gauge group $SU(3)_c \times SU(2)_L \times U(1)_Y $, along with the addition of three RHNs. To prevent the flavor-changing neutral current (FCNC), $Z_2$ symmetry is imposed under which the new scalar doublet $\eta$ is charged, whereas other 
particles are neutral. Gauge symmetries do not detect the 
flavor structure of the model. To have a 
predictive flavor structure of the lepton sector, we impose
$A_4$ flavor symmetry. So, two more SM scalar singlets 
$\chi$ and $\chi^\prime$ are added which transform as triplets under $A_4$. 
The SM doublet leptons $L_{e,\mu,\tau}$ transform as $1$, $1^{\prime}$ and $1^{\prime\prime}$, respectively, under $A_4$. The three corresponding charged leptons $e_R$, $\mu_R$, and $\tau_R$ transform in a similar fashion as the lepton doublets under $A_4$. Right-handed Majorana neutrinos ($N_R$), IHD $\eta$, and singlet scalars ($\chi$ and $\chi^\prime$) are triplets while the SM scalar doublet $\Phi$ is a singlet under $A_4$. 
This $Z_2$ symmetry prohibits the Yukawa term involving lepton doublets $L_{e,\mu,\tau}$, Majorana neutrinos $N_R$, and SM Higgs field $\Phi$; as a result, the Dirac mass term is absent. 
We also consider the existence of the dark sector composed of a SM singlet Dirac fermion $\Psi$ and a real singlet field $S$. This dark sector is differentiated by a $Z_2^{\prime}$ symmetry under which only these two fields are odd, while others are even. Similarly $Z_3$ is added to differentiate the $\chi$ and $\chi^{\prime}$. 
All the field contents and their different charge assignments under $SU(2)_L \times U(1)_Y\times A_4 \times Z_2 \times Z^{\prime}_2 \times Z_3$ are shown in Table (\ref{tab:fields}).
\begin{table}[htbp!]
\centering
\begin{tabular}{|c c c c c c c c|c c |}
\hline
Field & $L_e, L_{\mu}, L_{\tau}$ & $e_R, \mu_R, \tau_R$ & $N_R$ & $\chi$&$\chi^{\prime}$ & $\Phi$ & $\eta$ & $S$ & $\Psi$ \\
\hline
$A_4$ & $1, 1', 1''$ & $1, 1', 1''$ & $3$ & $3$&$3$ & $1$ & $1 $ & $1$ & $1$ \\
$Z_2$ & $+$ & $+$ & $-$ & $+$&$+$ & $+$ & $-$ & $-$ & $+$ \\
$Z_2'$ & $+$ & $+$ & $+$ & $+$&$+$ & $+$ & $+$ & $-$ & $-$ \\
$Z_3$ & $\omega$ & $\omega$ & $1$ & $1$ &$\omega^2$& $1$ & $1$ & $\omega$ & $1$ \\
$SU(2)_L \times U(1)_Y$ & $(2,-1)$ & $(1,-2)$ & $(1,0)$ & $(1,0)$ &$(1,0)$& $(2,1)$ & $(2,1)$ & $(1,0)$ & $(1,0)$ \\

\hline
\end{tabular}
\caption{Representation of fields under $  SU(2)_L \times U(1)_Y \times A_4 \times Z_2 \times Z_2'\times Z_3$.}
\label{tab:fields}
\end{table}
The scalar potential of the model is given by,
\begin{eqnarray}
\nonumber
 V &= & V(\Phi)+V(\eta)+ V(S) +V(\chi)+V(\chi')+V(\Phi, \chi) +V(\Phi, \chi')
 +V(\Phi,\eta) \\
 &+& V(\Phi,S)+V(\eta, \chi)+V(\eta, \chi') +V(\eta,S)+V(\chi,\chi') 
 +  V(\chi, S)+ V(\chi', S),
 \label{eq:pot}
\end{eqnarray}
where the individual expressions are given in the Appendix (\ref{app:potential}). Since we impose the CP invariance at the Lagrangian level, the coefficients of all the field operators are assumed to be real. The $A_4$ multiplication rules are given in the Appendix (\ref{app:mult-rule}).

Our starting point is an effective Lagrangian with $A_4 \times Z_2\times Z_2'\times Z_3$ symmetry. Below a cutoff scale $\Lambda$, the physics is expressed in terms of higher-dimensional operators. The scale 
at which $A_4$ is broken is above the electroweak scale and is assumed to be lower than the cutoff scale $\Lambda$ but still close to it.
With the field contents and symmetries specified in Table \ref{tab:fields}, the leptonic  part of the Lagrangian is given by,
\begin{eqnarray}
   -{\cal L} \supset  -{\cal L_{\rm vis}}- {\cal L_{D}},  
\end{eqnarray}
where ${\cal L_{\rm vis}}$ and ${\cal L_{D}}$ represent the visible and dark sector
Lagrangian, respectively. They can be explicitly written as,
\begin{eqnarray}\label{eq:yukawa lag}
 -{\cal L_{\rm vis}} &=& \frac{{y^\prime}^{\nu}_{1}}{\Lambda}y\bar{L}_{e}( \bar{\chi^{\prime}} N_{R})_{{\bf 1}}\tilde{\eta}+\frac{{y^\prime}^{\nu}_{2}}{\Lambda}\bar{L}_{\mu}(\bar{\chi^{\prime}}N_{R})_{{\bf 1}'} \tilde{\eta}+\frac{{y^\prime}^{\nu}_{3}}{\Lambda}\bar{L}_{\tau}(\bar{\chi^{\prime}}N_{R})_{{\bf 1}''}\tilde{\eta}
 +\frac{M}{2}(\overline{N^{c}_{R}}N_{R})_{{\bf 1}}
+ \frac{\lambda_{\chi}}{2}(\overline{N^{c}_{R}}N_{R})_{{\bf 3}_{s}} \chi
\nonumber\\
&+& y_{e}\bar{L}_{e}\Phi~e_{R}+ y_{\mu}\bar{L}_{\mu}\Phi~\mu_{R} 
+ y_{\tau}\bar{L}_{\tau}\Phi~ \tau_{R} +{\rm h.c.},
 \label{eq:lagrangian-vis}
 \end{eqnarray}
 \begin{eqnarray}
   -{\cal L_{D}} =  m_{\Psi}\bar{\Psi}\Psi + \frac{\lambda_{D}}{\Lambda}(\bar{N_R}\chi^{\prime})_1 S \Psi +  {\rm h.c}~.
   \label{eq:dark-L}
 \end{eqnarray}
The contribution of the higher-order $(d > 5)$ terms can be safely ignored as these terms will be suppressed due to the high cutoff scale $\Lambda$.
 Here, $\lambda_{D}$ is the coupling among the dark sector and the RHNs. The Dirac nature of the DM mass ($m_{\Psi}\bar{\Psi}\Psi$) preserves the lepton number. All the coupling constants are real because of the CP symmetry. Heavy neutrinos $N_R$ acquire bare mass M, as well as a mass induced by the vacuum of the $A_4$ triplet scalar $\chi$. In the above Lagrangian, we also include the masses for dark sector Dirac fermions as $m_{\Psi}$.
\subsection{Spontaneous CP Violation from the breaking of $A_4$ symmetry}
 {\label{sec:scpv}}
 The Lagrangian in equations (\ref{eq:lagrangian-vis}) and (\ref{eq:dark-L}) is CP symmetric. The CP is spontaneously violated due to 
 the complex VEV of the $A_4$ triplet scalar $\chi$. At the first stage, 
  $\chi$ and $\chi^{\prime}$ develop non-zero VEVs through which flavor symmetry $A_4$ and $Z_3$ will be broken. The VEVs are as given as 
 \begin{eqnarray}
 \langle\eta\rangle =0,~ \langle\chi\rangle = v_{\chi}e^{i\phi}(1,0,0),~ \langle\chi^{\prime}\rangle = v_{\chi}^{\prime}(1,1,1).
 \label{eq:vevs}
 \end{eqnarray}
The choice of the above particular VEV is discussed in the Appendix (\ref{subsec:minimize}). Since the scale of $A_4$ symmetry breaking is at a very high scale (much before electroweak symmetry breaking), the VEVs of $\chi$ and 
$\chi^\prime$ are very heavy and are decoupled from the electroweak scale. At this stage, the SM Higgs scalar VEV is zero. However, in the appendix(\ref{subsec:minimize})
we have considered nonzero VEVs for all scalar fields of the model
except IHD, to analyze the full scalar spectrum of the model.
In this section, we deal only with the CP-breaking minima, which arise from
$A_4$ symmetry breaking; we take the SM Higgs scalar VEV to be zero.
 
In the above, we have taken the VEV of $\chi$ to be complex through a phase $\phi$ as a result of which the CP symmetry will be broken spontaneously. The analysis for minimization of the scalar field has been carried out in the appendix (\ref{subsec:minimize}).  Minimizing the scalar potential with different values
of the CP-violating phase $\phi$ and the corresponding minima leads to three choices, given as\\
 {\bf (i)} for $\phi=0, \pm\pi$
 \begin{equation}
V_0 = 3 m_{\chi}^2 \, v_{\chi^\prime}^2
+ 12 \xi_1^{\chi^\prime} v_{\chi^\prime}^3
+ 12 v_{\chi^\prime}^4 \lambda_5^{\chi^\prime} 
+ \frac{3}{8} \frac{\Big[ m_{\chi}^2
+ v_{\chi^\prime}^2 \big( 3 \lambda_1^{\chi\chi^\prime}
+ 2 \lambda_2^{\chi\chi^\prime}
+ 2\lambda_3^{\chi\chi^\prime} \big)
+ 2 \mu_{\chi}^2 \Big]^2}{\lambda_1^{\chi} + \lambda_2^{\chi} + \tilde{\lambda}_2^{\chi}},
\end{equation}
{\bf (ii)} for $\phi=\pm\pi/2$
 \begin{equation}
V_0= 3 m_{\chi}^2 \, v_{\chi^\prime}^2
+ 12 \xi_1^{\chi^\prime} v_{\chi^\prime}^3
+ 12 v_{\chi^\prime}^4 \lambda_5^{\chi^\prime}
 + \frac{3}{8} \frac{\Big[ m_{\chi}^2
+ v_{\chi^\prime}^2 \big( 3 \lambda_1^{\chi\chi^\prime}
+ 2 \lambda_2^{\chi\chi^\prime}
+2 \lambda_3^{\chi\chi^\prime} \big)
- 2 \mu_{\chi}^2 \Big]^2}{\lambda_1^{\chi} + \lambda_2^{\chi} - \tilde{\lambda}_2^{\chi}},
\label{VP1}
\end{equation}
{\bf (iii)} for $\cos2\phi=-\frac{\mu^{2}_{\chi}+v^{2}_{\chi}\tilde{\lambda}^{\chi}_{2}}{4v^{2}_{\chi}(\lambda^{\chi}_{1}+\lambda^{\chi}_{2})}$
\begin{equation}
V_0 = 3 m_{\chi}^2 \, v_{\chi^\prime}^2
+ 12 \xi_1^{\chi^\prime} v_{\chi^\prime}^3
+ 12 v_{\chi^\prime}^4 \lambda_5^{\chi^\prime}
+ \frac{1}{8} \frac{\Big[ m_{\chi}^2
+ v_{\chi^\prime}^2 \big( 3 \lambda_1^{\chi\chi^\prime}
+ 2 \lambda_2^{\chi\chi^\prime}
+2 \lambda_3^{\chi\chi^\prime} \big) \Big]^2}{\lambda_1^{\chi} + \lambda_2^{\chi}}.
\end{equation}
The first case (i) corresponds to CP conservation of the vacuum configuration, whereas the cases (ii) and (iii) clearly show the breaking of CP symmetry.
In all of the above cases, we can obtain global minima by choosing the appropriate choice of parameters involved in $V_0$. For, $m_{\chi}^2<0$, $\mu_{\chi}^2 < 0$, $\lambda^{\chi}_{1,2}, \lambda^{\chi^\prime}_5,\xi^{\chi^\prime}_1<0$, and $\tilde{\lambda}^{\chi}_{2}>0$, case (iii) leads to the absolute minimum of the potential. As $V_0$ is calculated for spontaneous CP violation with nonzero $\phi$, our model with spontaneous CP violation is justified with this global minimum condition.
\subsection{Lepton mass matrices}
\label{subsec-lepton}
After the first stage of $A_4$ symmetry breaking, the visible sector Lagrangian in Eq.(\ref{eq:lagrangian-vis})
takes the form,
 \begin{eqnarray}
 -{\cal L}_{\rm vis} &=& y^{\nu}_{1}\bar{L}_{e}\tilde{\eta}(N_{R_1}+N_{R_2}+N_{R_3})+y^{\nu}_{2}\bar{L}_{\mu}\tilde{\eta}(N_{R_1}+\omega N_{R_2}+\omega^2N_{R_3})\nonumber\\
 &+&y^{\nu}_{3}\bar{L}_{\tau}\tilde{\eta}(N_{R_1}+\omega^2N_{R_2}+\omega N_{R_3})\nonumber\\
 &+&\frac{M}{2}(\overline{N^{c}_{R_1}}N_{R_1} + \overline{N^{c}_{R_2}}N_{R_2} + \overline{N^{c}_{R_3}}N_{R_3} )+\frac{\lambda^s_{\chi}v_{\chi}e^{i\phi}}{2}(\overline{N^{c}_{R_2}}N_{R_3} + \overline{N^{c}_{R_3}}N_{R_2}) \nonumber\\
 &+& y_{e}\bar{L}_{e}\Phi~e_{R}+y_{\mu}\bar{L}_{\mu}\Phi~\mu_{R}+y_{\tau}\bar{L}_{\tau}\Phi~ \tau_{R} + {\rm h.c.}
 \label{lagrangian-yuk}
 \end{eqnarray}
 and that corresponding to the dark sector is given by,
\begin{eqnarray}
 -{\cal L}_{D} = m_{\Psi}\bar{\Psi}\Psi + \lambda_{d}(\overline{N}_{R1} S\Psi+\overline{N}_{R2} S\Psi+\overline{N}_{R3} S\Psi) +{\rm h.c},
\end{eqnarray}
where the effective coupling connecting the DM sector and the SM sector becomes $\lambda_{d}=\lambda_{D}v^{\prime}_{\chi}/\Lambda$ and $y^{\nu}_i = {y^\prime}^{\nu}_i v_{\chi'}/\Lambda$, $\Lambda$ being the cut-off scale of our model.  The massive radial modes are settled to the minimum and are ignored below the symmetry-breaking scale. In the next stage corresponding to the electroweak symmetry breaking, 
the SM Higgs scalar $\Phi$ develops a non-zero VEV as  
\begin{equation}
   \langle\Phi\rangle =
   \begin{pmatrix}
    0\\
     v_{\Phi} e^{i\theta}
   \end{pmatrix}~,\quad
   \end{equation} 
where $v_{\Phi}= 174$ GeV. Here the phase $\theta$ is not a
physical observable, so we set it to zero. The mass matrix for the charged leptons is generated, and the Lagrangian reduces to the form,
 \begin{equation}
 -{\mathcal{L}_{\rm vis}} \supset \bar{L_l} m_l L_R + \bar{L_l} Y_\nu
 \tilde{\eta} N_R + \frac{1}{2} \bar{N_R^c} M_R N_R 
 + {\rm h.c.},
 \label{eq:lepton-Langr}
\end{equation}
where
\begin{eqnarray}
\qquad m_{l}=v_{\Phi}{\left(\begin{array}{ccc}
 y_{e} & 0 & 0 \\
 0 & y_{\mu} & 0 \\
 0 & 0 &  y_{\tau}
 \end{array}\right)},~~
  M_{R}={\left(\begin{array}{ccc}
 M &  0 &  0 \\
 0 &  M &  \lambda^{s}_{\chi}v_{\chi} e^{i\phi} \\
 0 &  \lambda^{s}_{\chi}v_{\chi} e^{i\phi} &  M
 \end{array}\right)}.
 \label{eq:majorana and charged lepton mass-matrix}
 \end{eqnarray}
In addition to this, one can easily see that the neutrino Yukawa matrix is given as follows:
\begin{equation}
 Y_{\nu}=\sqrt{3}{\left(\begin{array}{ccc}
 y^{\nu}_{1} &  0 &  0 \\
 0 & y^{\nu}_{2} & 0 \\
 0 & 0 & y^{\nu}_{3}
 \end{array}\right)}U^{\dag}_{\omega}~,\qquad
 U_{\omega}=\frac{1}{\sqrt{3}}{\left(\begin{array}{ccc}
 1 &  1 &  1 \\
 1 &  \omega^{2} &  \omega \\
 1 &  \omega &  \omega^{2}
 \end{array}\right)}~.
 \label{yuk-matrix}
 \end{equation}
The charged lepton mass matrix turns out to be diagonal. It is useful to work in a basis 
where the heavy Majorana neutrino mass matrix is also diagonal. Rotating the basis with
the help of a unitary matrix $U_R$,
 \begin{equation}
 N_{R}\rightarrow U^{\dag}_{R}N_{R}~,
 \label{basis}
 \end{equation}
 the right-handed Majorana mass matrix $M_R$ becomes a diagonal matrix $\tilde{M_R}$ and is given by \cite{Ahn:2013mva}
\begin{equation}
 \tilde{M}_{R}=U^{T}_{R}M_{R}U_{R}=MU^{T}_{R}{\left(\begin{array}{ccc}
 1 &  0 &  0 \\
 0 &  1 & \kappa e^{i\phi} \\
 0 &  \kappa e^{i\phi} &  1
 \end{array}\right)}U_{R}={\left(\begin{array}{ccc}
 aM & 0 & 0 \\
 0 & M & 0 \\
 0 & 0 & bM
 \end{array}\right)}~,
 \label{MR1}
 \end{equation}
where $\kappa=\lambda^{s}_{\chi}v_{\chi}/M$. The unremovable phase from the Majorana mass matrix is the origin of CP violation in low- and high-energy regimes.
It can be found that $a=\sqrt{1+\kappa^{2}+2\kappa\cos\phi}$, $b=\sqrt{1+\kappa^{2}-2\kappa\cos\phi}$, and the diagonalizing matrix
\begin{equation}
  U_{R} = \frac{1}{\sqrt{2}} {\left(\begin{array}{ccc}
  0  &  \sqrt{2}  &  0 \\
  1 &  0  &  -1 \\
  1 &  0  &  1
  \end{array}\right)}{\left(\begin{array}{ccc}
  e^{i\frac{\psi_1}{2}}  &  0  &  0 \\
  0  &  1  &  0 \\
  0  &  0  &  e^{i\frac{\psi_2}{2}}
  \end{array}\right)}~,
  \label{UR}
\end{equation}
with the phases
\begin{equation}
 \psi_1 = \tan^{-1} \Big( \frac{-\kappa\sin\phi}{1+\kappa\cos\phi} \Big),~~~ \psi_2 = \tan^{-1} \Big( \frac{\kappa\sin\phi}{1-\kappa\cos\phi} \Big)~.
\label{alphs_beta}
\end{equation}
It can be observed that the phases $\psi_{1,2}$ can go to $0$ or $\pi$ with a real VEV of 
the flavon field $\chi$, {\it i.e.} for $\phi =0$. Due to the rotation, the neutrino Yukawa matrix 
$Y_{\nu}$ also gets redefined as,
\begin{equation}
 \tilde{Y}_{\nu} = Y_{\nu}U_{R}~
    = P_{\nu}^{\dag}~{\rm Diag}(|y^{\nu}_{1}|,|y^{\nu}_{2}|,|y^{\nu}_{3}|)U^{\dag}_{\omega}U_{R}~.
 \label{modified-yuk}
 \end{equation}

Now performing the basis rotation in the leptonic sector from weak to mass eigenstate,
\begin{equation}
 l_{L}\rightarrow P^{\ast}_{\nu}l_{L}~,\qquad l_{R}\rightarrow P^{\ast}_{\nu}l_{R}~,\qquad\nu_{L}\rightarrow U^{\dag}_{\nu}P^{\ast}_{\nu}\nu_{L}
 \label{basis-rot}
 \end{equation}
where $P_{l}$ and $P_{\nu}$ are phase matrices and $U_{\nu}$ is a diagonalizing matrix of light neutrino mass matrix. So, we obtain the lepton mixing matrix $U_{\rm PMNS}$ as
\begin{equation}
 U_{\rm PMNS}=P^{\ast}_{l}P_{\nu}U_{\nu}~.
 \end{equation}

The $U_{\rm PMNS}$ matrix can be written in terms of three mixing angles and three CP-odd phases as follows
\begin{equation}
  U_{\rm PMNS}={\left(\begin{array}{ccc}
   c_{13}c_{12} & c_{13}s_{12} & s_{13}e^{-i\delta_{CP}} \\
   -c_{23}s_{12}-s_{23}c_{12}s_{13}e^{i\delta_{CP}} & c_{23}c_{12}-s_{23}s_{12}s_{13}e^{i\delta_{CP}} & s_{23}c_{13}  \\
   s_{23}s_{12}-c_{23}c_{12}s_{13}e^{i\delta_{CP}} & -s_{23}c_{12}-c_{23}s_{12}s_{13}e^{i\delta_{CP}} & c_{23}c_{13}
   \end{array}\right)}Q_{\nu}~,
 \label{pmns}
 \end{equation}
where $s_{ij}= \sin\theta_{ij}$ and $c_{ij}\equiv \cos\theta_{ij}$, and $Q_{\nu}={\rm Diag}(e^{-i\psi_{1}/2},e^{-i\psi_{2}/2},1)$.

By virtue of an appropriate redefinition of the left-handed charged lepton fields, we can take the eigenvalues $y^{\nu}_1$, $y^{\nu}_2$, and $y^{\nu}_3$ to be real and positive. So the Yukawa matrix can be written as
 \begin{equation}
 Y_{\nu}=y^{\nu}_{3}\sqrt{3}{\left(\begin{array}{ccc}
 y_{1} &  0 &  0 \\
 0 & y_{2} & 0 \\
 0 & 0 & 1
 \end{array}\right)}U^{\dag}_{\omega},
  \label{new-yuk}
 \end{equation}
where $y_{1}=|y^{\nu}_{1}/y^{\nu}_{3}|, y_{2}=|y^{\nu}_{2}/y^{\nu}_{3}|$.

However, the underlying CP violation pertinent to leptogenesis arise from the neutrino Yukawa matrix $\widetilde{Y}_{\nu}=Y_{\nu}U_{R}$ and its combination $\tilde{Y}^{\dagger}_{\nu}\tilde{Y}_{\nu}$, which is given by

\begin{eqnarray}
   \tilde{Y}^{\dagger}_{\nu}\tilde{Y}_{\nu}=|y^{\nu}_{3}|^{2}\left(\begin{array}{ccc}
  \frac{1+4y^{2}_{1}+y^{2}_{2}}{2} & \frac{e^{-i\frac{\psi_{1}}{2}}}{\sqrt{2}}(2y^{2}_{1}-y^{2}_{2}-1) & \frac{i\sqrt{3}e^{i\frac{\psi_{21}}{2}}}{2}(y^{2}_{2}-1) \\
  \frac{e^{i\frac{\psi_{1}}{2}}}{\sqrt{2}}(2y^{2}_{1}-y^{2}_{2}-1) & 1+y^{2}_{1}+y^{2}_{2} & -i\sqrt{\frac{3}{2}}e^{i\frac{\psi_{2}}{2}}(y^{2}_{2}-1) \\
  -\frac{i\sqrt{3}e^{-i\frac{\psi_{21}}{2}}}{2}(y^{2}_{2}-1) & i\sqrt{\frac{3}{2}}e^{-i\frac{\psi_{2}}{2}}(y^{2}_{2}-1) & \frac{3}{2}(1+y^{2}_{2})
  \end{array}\right) ,
 \label{ynuynu}
 \end{eqnarray}
 where $\psi_{ij}\equiv\psi_{i}-\psi_{j}$. The non-zero value of 
 CP violation in two-sector leptogenesis requires the coefficients of the 
 $e^{\pm i \psi_{1,2}}$-terms in the off diagonal elements of the
 matrix in Eq.(\ref{ynuynu}) to be non-zero. In the limit $y_1^\nu =
 y_2^\nu=y_3^\nu$, the off-diagonal elements of $\tilde{Y}^{\dagger}_{\nu}\tilde{Y}_{\nu}$ vanish. So, a non-zero value of CP violation, both in a low-scale phenomenon
 like neutrino oscillation and a high-scale phenomena like two-sector leptogenesis requires hierarchical
 Dirac-type neutrino Yukawa couplings and a non-zero phase $\psi_{1,2}$ emerging
 from spontaneous CP symmetry breaking.
\begin{figure}[htbp!]
 \centering
 \includegraphics[width=0.4\linewidth,height=3.5cm]{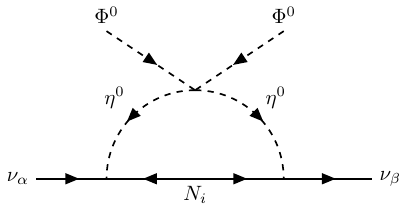}
 \caption{Diagram for light neutrino mass generation through one-loop.}
 \label{fig:loop}
\end{figure}
\subsection{Light neutrino mass matrix}
\label{sec:nu-matrix}
As the $Z_2$ symmetry is conserved even after electroweak symmetry breaking, the neutrino masses will be generated radiative mechanism.
With our model construction, the heavy RHNs do not couple to the SM Higgs scalar. So
it can not develop a Dirac mass term. But it can do so via its coupling to the IHD $\eta$ at one-loop level. 
In the present model, the light neutrino mass matrix can be generated via the radiative seesaw mechanism, where the RHNs and the IHD participate through a one-loop diagram as shown in Fig.(\ref{fig:loop}). After electroweak symmetry breaking, the light neutrino masses in the flavor basis, where the charged lepton mass matrix is real and diagonal, are written as \cite{Ahn:2013mva}
\begin{equation}
   (m_{\nu})_{\alpha\beta}=\sum_{i}\frac{\Delta m^{2}_{\eta_i}}{16\pi^{2}}\frac{(\tilde{Y}_{\nu})_{\alpha i}(\tilde{Y}_{\nu})_{\beta i}}{M_{i}}f\left(\frac{M^{2}_{i}}{\bar{m}^{2}_{\eta_i}}\right),~~\Delta m^{2}_{\eta_{i}} \ll  \bar{m}^{2}_{\eta_i}~,
   \label{nu-mass}
 \end{equation}
where
 \begin{equation}
   f(z_{i})=\frac{z_{i}}{1-z_{i}}\left[1+\frac{z_{i}\ln z_{i}}{1-z_{i}}\right]~,
 \label{loop-fun}
 \end{equation}
\begin{equation}
 \Delta m^{2}_{\eta_{i}}\equiv|m^{2}_{R_i}-m^{2}_{I_i}|=8v^{2}\lambda^{\Phi\eta}_{3}~,
\end{equation}
with $z_{i}=M^{2}_{i}/\bar{m}^{2}_{\eta_i}$ and $\bar{m}^{2}_{\eta_i}\equiv(m^{2}_{R_i}+m^{2}_{I_i})/2$.  Here, $m_{R} ~\rm{and}~ m_{I}$ are the mass of the real part ($h_1$) and imaginary part ($A_1$) of the neutral component of IHD. The explicit expression for associated mass parameters is presented in Appendix (\ref{higgs-mass}).

 With $\tilde{M}_{R}={\rm Diag}(M_{r1},M_{r2},M_{r3})$ and $M_{ri}= M_{i}f^{-1}(z_{i})$, the above formula can be expressed as
 \begin{eqnarray}
  m_{\nu} &=& \frac{v^2_{\Phi}\lambda^{\Phi\eta}_{3}}{4\pi^{2}}\tilde{Y}_{\nu}\tilde{M}^{-1}_{R}\tilde{Y}^{T}_{\nu} \\
  &=& U_{\rm PMNS}~{\rm Diag}(m_{1},m_{2},m_{3}) U^{T}_{\rm PMNS} \label{nu-massmatrix}\\
  &=& m_{0}{\left(\begin{array}{ccc}
  Ay^{2}_{1} & By_{1}y_{2} & By_{1} \\
  By_{1}y_{2} & Dy^{2}_{2} & Gy_{2}  \\
  By_{1} & Gy_{2} & D
 \end{array}\right)}~,
 \label{radseesaw}
 \end{eqnarray}
where $m_{i}(i=1,2,3)$ are the light neutrino mass eigenvalues, $y_{1(2)}=y^{\nu}_{1(2)}/y^{\nu}_3$, and
 \begin{eqnarray}
 A&=&f(z_{2})+\frac{2e^{i\psi_{1}}f(z_{1})}{a}~, B=f(z_{2})-\frac{e^{i\psi_{1}}f(z_{1})}{a}~,\nonumber\\
  D&=&f(z_{2})+\frac{e^{i\psi_{1}}f(z_{1})}{2a}-\frac{3e^{i\psi_{2}}f(z_{3})}{2b}~,
  m_{0}= \frac{v^2_{\Phi}|y^{\nu}_{3}|^{2}\lambda^{\Phi\eta}_{3}}{4\pi^{2}M}~,\nonumber\\
  G&=&f(z_{2})+\frac{e^{i\psi_{1}}f(z_{1})}{2a}+\frac{3e^{i\psi_{2}}f(z_{3})}{2b}~.
 \label{entries}
 \end{eqnarray}
 The matrix configuration depends on the values of $y_1$ and $y_2$. In the next sections, we study the baryon asymmetry along with the DM asymmetry for our model. Then, we show numerically that all these things are correlated. 
\section{Asymmetric dark matter and BAU in two-sector leptogenesis}
\label{adm-bau}
In this section, we discuss the framework of two-sector leptogenesis in the model
given in section (\ref{model}), where the RHNs relate the visible and the dark sectors.
The part of the Lagrangian responsible for two-sector leptogenesis reads as
\begin{eqnarray}
    -{\cal L} \supset \bar{L_l} Y_\nu
 \tilde{\eta} N_R + \frac{1}{2} \bar{N_R^c} M_R N_R + m_{\Psi}\bar{\Psi}\Psi + \lambda_{d}(\overline{N}_{R1} S\Psi+\overline{N}_{R2} S\Psi+\overline{N}_{R3} S\Psi) +h.c~.
\end{eqnarray}

\begin{figure}[htbp!]
    \centering
    \subcaptionbox{\label{fig:sub14}}{\includegraphics[width=0.3\linewidth,height=3cm]{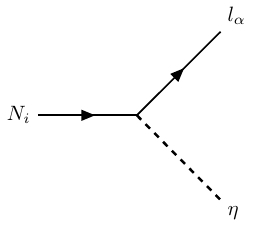}}
    \subcaptionbox{\label{fig:sub15}}{\includegraphics[width=0.3\linewidth,height=3cm]{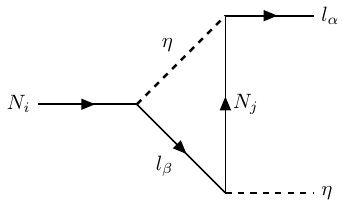}}\\
    \subcaptionbox{\label{fig:sub16}}{\includegraphics[width=0.3\linewidth,height=3cm]{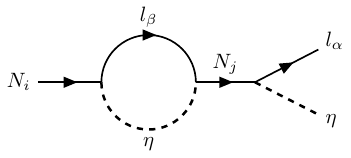}}
    \subcaptionbox{\label{fig:sub17}}{\includegraphics[width=0.3\linewidth,height=3cm]{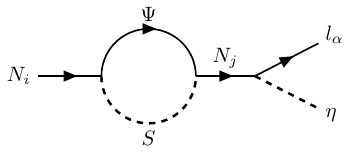}}
     \caption{Diagrams generating lepton asymmetry in the visible sector.}
    \label{fig:set3}
\end{figure}

\begin{figure}[htbp!]
    \centering
    \subcaptionbox{\label{fig:sub18}}{\includegraphics[width=0.3\linewidth,height=3cm]{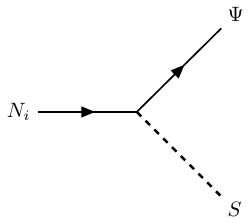}}
    \subcaptionbox{\label{fig:sub19}}{\includegraphics[width=0.3\linewidth,height=3cm]{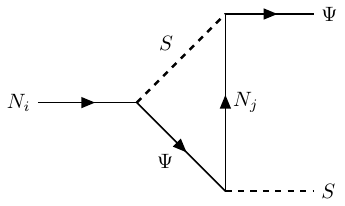}}\\
    \subcaptionbox{\label{fig:sub20}}{\includegraphics[width=0.3\linewidth,height=3cm]{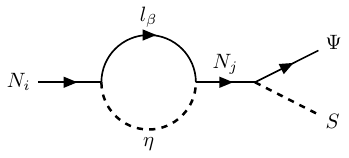}}
    \subcaptionbox{\label{fig:sub21}}{\includegraphics[width=0.3\linewidth,height=3cm]{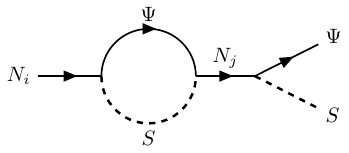}}
     \caption{Diagrams generating asymmetry in dark sector.}
    \label{fig:set3.1}
\end{figure}
To realize the asymmetric dark matter scenario in our scotogenic model, we consider the lightest RH neutrino $N_1$ as a mediator between the two sectors, i.e., the IHD extended SM sector and the dark sector. It can decay into the IHD and the SM lepton doublet through the Yukawa coupling $Y_\nu$ and into the scalar $S$ and fermion $\Psi$ through the Yukawa coupling $\lambda_{d}$. It is worth recalling that the core purpose of our study is to supplement the deficit in the DM relic density corresponding to the IHD desert. In this context, once the symmetric component of DM particle $\Psi$ is annihilated away, the asymmetries in its number densities determine the relic density of DM. Owing to the lepton number associated with the DM particle $\Psi$, lepton asymmetries are produced in both sectors. 

The asymmetries produced in both sectors from the decay of $N_1$ originate from the interference between the tree-level and one-loop diagrams as shown in figures (\ref{fig:set3}) and (\ref{fig:set3.1}), and are given by \cite{Biswas:2018sib, Covi:1996wh, Siyeon:2003bj},
\begin{eqnarray}
\epsilon_L &=& \frac{\sum_{\alpha}\left[\Gamma(N_1 \rightarrow L_{\alpha} + \eta) - \Gamma(N_1 \rightarrow \bar{L_{\alpha}} + \eta^{\ast})\right]}{\Gamma_1}~,\\
&\simeq& \frac{1}{16\pi} \sum_{k=2,3} \frac{\rm{Im}\left[3((\tilde{Y}^{\dagger}_{\nu}\tilde{Y}_{\nu})^{\ast}_{1k})^2 + 2{\lambda}^{2}_{d}(\tilde{Y}^{\dagger}_{\nu}\tilde{Y}_{\nu})^{\ast}_{1k})\right]}{\left[(\tilde{Y}^{\dagger}_{\nu}\tilde{Y}_{\nu})_{11} + |\lambda_{d}|^2 \right]
}~\left(\frac{M_1}{M_k}\right)~,
 \label{epsion-l}
\end{eqnarray}
and
\begin{eqnarray}
 \epsilon_{\Psi} &=& \frac{\sum_{\alpha}\left[\Gamma(N_1 \rightarrow \Psi + S) - \Gamma(N_1 \rightarrow \bar{\Psi} + S^{\ast})\right]}{\Gamma_1}~, \\
 &\simeq& \frac{1}{16\pi} \sum_{k=2,3} \frac{\rm{Im}\left[ 2{\lambda}^ {2}_{d}(\tilde{Y}^{\dagger}_{\nu}\tilde{Y}_{\nu})^{\ast}_{1k}) + 3({\lambda}_{d})^4\right]}{\left[(\tilde{Y}^{\dagger}_{\nu}\tilde{Y}_{\nu})_{11} + |\lambda_{d}|^2 \right]
}~\left(\frac{M_1}{M_k}\right)~,
 \label{epsilon-psi}
\end{eqnarray}
where
\begin{equation}
 \Gamma_1 = \frac{M_1}{8\pi}\left((\tilde{Y}^{\dagger}_{\nu}\tilde{Y}_{\nu})_{11} +|\lambda_{d}|^2\right)~,
 \label{eq-gamma1}
\end{equation}
is the total decay width of $N_1$.
Now using Eq.(\ref{ynuynu}) in equations  (\ref{epsion-l}) and (\ref{epsilon-psi}), the asymmetries can be written as follows,
\begin{align}
\epsilon_L &= \frac{1}{16\pi C}\Bigg[
\left(\tfrac{3}{2}{y^\nu_3}^4(2y_1^2-y_2^2-1)^2\sin\psi_1
- \sqrt{2}\,\lambda_d^2 {y^\nu_3}^2(2y_1^2-y_2^2-1)\sin\frac{\psi_1}{2}\right)\frac{M_1}{M_2} \nonumber\\
&\qquad\qquad
+ \left(\tfrac{9}{4}{y^\nu_3}^4(y_2^2-1)^2\sin\psi_{21}
+ \sqrt{3}\,\lambda_d^2 {y^\nu_3}^2(y_2^2-1)\cos\frac{\psi_{21}}{2}\right)\frac{M_1}{M_3}
\Bigg], \label{analytic-epsilon-l} \\[0.6em]
\epsilon_{\psi} &= \frac{1}{16\pi C} \Bigg[
\sqrt{2} \lambda_d^2 {y^\nu_3}^2(2y_1^2-y_2^2-1)\sin\frac{\psi_1}{2} \frac{M_1}{M_2}
- \sqrt{3} \lambda_d^2 {y^\nu_3}^2(y_2^2-1)\cos\frac{\psi_{21}}{2} \frac{M_1}{M_3}
\Bigg],\label{analytic-epsilon-psi}
\end{align}
where $C=\left( \frac{1 + 4 y_1^2 + y_2^2}{2} \, {y^{\nu}_{3}}^2 + \lambda_{d}^2 \right)$.
It can be observed that
for a hierarchical mass limit among RHNs, $M_1\ll M_2 \ll M_3$, the second terms in right-hand-side of equations (\ref{analytic-epsilon-l}) and (\ref{analytic-epsilon-psi}) ($\propto
\frac{M_1}{M_3}$) are subdominant as compared to the first one. In the first term 
($\propto \frac{M_1}{M_2}$), the dependence on the CP-violating phase $\psi_1$ can be 
observed. If $\psi_1 \rightarrow 0$, both $\epsilon_{L}, \epsilon_{\psi}  \rightarrow 0$.
So, the phase $\psi_1$ acts as a common link between the visible and dark sectors as far as the generation of leptonic CP asymmetry is concerned. 
So even if $\lambda_{d}$ can be taken to be real, the CP asymmetry is nonzero 
due to the non-zero value of phase $\psi_1$. The relevant parameters required for the calculation 
of $\epsilon_{l}$ and $\epsilon_{\psi}$ can be determined from neutrino oscillation data,
which is discussed in the next section.

After calculating the lepton and dark matter asymmetries ($\epsilon_L$ and $\epsilon_{\Psi}$), their evolution can be studied by using the Boltzmann equations. Denoting the asymmetry abundances as $Y_{\Delta L}=Y_L-Y_{\bar L}$ and $Y_{\Delta\Psi}= Y_{\Psi}-Y_{\bar {\Psi}}$ for the visible sector and dark sector, respectively, the Boltzmann equations can be written as
\begin{eqnarray}
\frac{d Y_{N_1}}{dz}&=& -z\frac{\Gamma_1}{H_1}
\frac{K_1(z)}{K_2(z)}\left(Y_{N_1}-Y_{N_1}^{\rm eq}\right)~,
\label{eq:BE}
\\
\frac{d Y_{\Delta x}}{dz} &=&   \frac{\Gamma_1}{H_1}
\left (\epsilon_x  z \frac{K_1(z)}{K_2(z)}(Y_{N_1}- Y_{N_1}^{eq})  -  Br_x
\frac{z^3 K_1(z)}{4} Y_{\Delta x}  \right)~, (x=L,\Psi).
\label{eq:BE1}
\end{eqnarray}
Here $Y_{j}=\frac{n_j}{s}$ is the abundance yield, $n_j$ represent the respective number 
densities of the particle j ($j=N_1, x$ with $x = l,\Psi$), $s=\frac{2\pi^2}{45}  g_{*}  T^3$ is the entropy density at some temperature $T$ and $g_{*}=106.75$ is the effective number of relativistic degrees of freedom. $Y_{N_1}^{\rm eq}$ represents the equilibrium abundance of $N_1$ and 
$z$ is given by $z = \frac{M_1}{T}$.
$H_1 =  1.66 \,\sqrt{g_{*}} \,\frac{T^2}{M_{\rm Pl}}$ is the Hubble parameter at $T = M_1$ and $M_{\rm Pl} \;=\; 1.22 \times 10^{19}\ \text{GeV}$. $K_i(z),~i=1,2$ are the modified Bessel 
functions of first and second kinds, respectively, and $Br_{x}$ denotes the branching
ratios of $N_1$ decay into the two sectors.
Eq.(\ref{eq:BE}) describes the evolution of $N_1$ abundance due to its decay and inverse decay. 
Eq.(\ref{eq:BE1}) describes the evolution of asymmetries in both sectors. The first term within the first bracket of Eq.(\ref{eq:BE1}), proportional to $\epsilon_{L (\Psi)}$, corresponds to the asymmetry $Y_{\Delta L ({\Delta \Psi})}$ production when $N_1$ departs from the thermal equilibrium and the second term, proportional to $Br_{L(\Psi)}$, represent the washout of asymmetries due to inverse decay of $N_1$.
The factor $\Gamma_1/H_1$ not only defines the strength of those interactions, but also determines the out-of-equilibrium behavior of $N_1$ (necessary to comply with Sakharov's conditions).

Depending on the decay width and mass of RHNs, the asymmetry production in both the sectors can be categorized into two regimes: one is the narrow width approximation, where $\Gamma_1 \ll  M_1$, and the other one is the strong washout regime where $\Gamma_1 \simeq M_1$ \cite{Falkowski:2011xh}. By considering the narrow-width approximation,  the source of washout effects can be neglected. As can be seen from Eq.(\ref{eq:BE1}), we have only considered the inverse decay as the dominant source of washout and circumvented the other washout terms like $L \Phi \leftrightarrow \bar{L} \Phi^{\dagger},~\bar{\Psi} S \leftrightarrow \Psi S$, mediated by $N_1$. The narrow-width approximation also prevents transfer of asymmetries between the visible sector and dark sector by restricting the $2 \leftrightarrow 2$ transfer terms ($L \Phi \leftrightarrow \bar{\Psi} S,~L \Phi \leftrightarrow \Psi S, \bar{L} \Phi^{\dagger} \leftrightarrow \bar{\Psi} S, \bar{L} \Phi^{\dagger} \leftrightarrow \Psi S$).

By solving the relevant Boltzmann equations, we find the final yields of 
lepton asymmetry in the visible and dark sectors. Further, a part of the lepton asymmetry is converted into the baryon asymmetry via the Sphaleron process. However, Sphaleron does not interact with the dark sector, so the dark sector asymmetry will not be converted. After the Sphaleron transition, the net baryon asymmetry is given by \cite{Davidson:2008bu}
\begin{equation}
 {\color{red}}Y_{\Delta B} = \frac{8}{23} Y_{\Delta L}. 
 \label{sphaleron}
\end{equation}
Also, in order to get the observed baryon asymmetry of the universe $Y_{\Delta B}=(8.24-9.38)\times10^{-11}$, $Y_{\Delta L}$ has to be within the range $(2.37-2.70)\times10^{-10}$ \cite{ParticleDataGroup:2018ovx}. Denoting $Y_{\Delta L}$ and $Y_{\Delta\Psi}$ as the final abundance yield at present-day temperature, the relic abundance of dark matter will be estimated as \cite{Edsjo:1997bg},
\begin{equation}
 \Omega_{\Psi}h^2= 2.755 \times 10^8 \left(\frac{m_{\Psi}}{{\rm{GeV}}}\right)Y_{\Delta \Psi}.
 \label{dark-relic}
\end{equation}
In the next section we present the numerical results of the analysis.
\section{Result}
\label{result}
In this section, one of our intentions is to explore the mass range for the dark matter candidate $\Psi$ in our model corresponding to the achievable dark matter relic density using Eq.(\ref{dark-relic}). For this we need to obtain the abundance yields for the both sector asymmetries i.e., $Y_{\Delta L~(\Delta \Psi)}$ by solving the Boltzmann equations (\ref{eq:BE}) and (\ref{eq:BE1}), which in turn depends on the value of CP asymmetry $\epsilon_{L(\Psi)}$ and the branching fraction $Br_{L(\Psi)}$. The evaluation of the asymmetries using equations (\ref{epsion-l}) and (\ref{epsilon-psi}), require the inputs from $\tilde{Y}_{\nu}$ matrix, $\lambda_{d}$ and $M_1$. Since we have performed a basis rotation, it is reasonable to adopt $\tilde{Y}_{\nu}$ instead of $Y_{\nu}$. As discussed in section (\ref{model}), we can find the $\tilde{Y}_{\nu}$ matrix by putting Eq.(\ref{new-yuk}) into Eq.(\ref{modified-yuk}). Similarly, the $M_1$ value can be taken from the diagonalised RHN mass matrix shown in Eq.(\ref{MR1}), according to which we can establish the hierarchy among the RH neutrinos, i.e., $M_1 < M_2 < M_3$ by imposing the condition $a<1 ~\rm{and}~ b>1$. In the course of our approach, the parameters $y^{\nu}_{3}, M, y_{1}, y_{2}, \kappa, \phi, \lambda^{\Phi\eta}_{3}, z_{1}, z_{2}, z_{3}, \lambda_{d}$ emerge and their values are subjected to the constraints arising from light neutrino mass matrix parameters, which are embedded in Eq.(\ref{nu-massmatrix}).  But our model parameters appear explicitly in Eq.(\ref{radseesaw}). So, first we reconstruct the light neutrino mass matrix using Eq.(\ref{nu-massmatrix})and the neutrino oscillation data shown in Table (\ref{osci-parameter}) for NH only. Fig.(\ref{fig:light-nu-range}) in Appendix \ref{light-nu-mass} depicts the possible range of each element of the neutrino mass matrix with an upper limit $\sim 10^{-11}$~GeV. This offers a means to constrain the respective matrix elements of $m_\nu$ shown in Eq.(\ref{radseesaw}). 
\begin{table}[h]
\centering
\renewcommand{\arraystretch}{1.3}
\begin{tabular}{|c | c|}
\hline
\textbf{Parameter} & \textbf{3$\sigma$ range} \\
\hline
$\theta_{12}/^\circ$        & $31.63 \to 35.95$ \\
$\theta_{23}/^\circ$        & $41.0 \to 50.5$   \\
$\theta_{13}/^\circ$        & $8.18 \to 8.87$   \\
$\delta_{\text{CP}}/^\circ$ & $96 \to 422$      \\
$\Delta m_{21}^2/10^{-5}\,\text{eV}^2$ & $6.92 \to 8.05$ \\
$\Delta m_{31}^2/10^{-3}\,\text{eV}^2$ & $+2.463 \to +2.606$ \\
\hline
\end{tabular}
\caption{Value of neutrino mixing angles and mass-squared differences considering Normal ordering of neutrino mass eigen states in $3 \sigma$ range taken from Ref.\cite{Esteban:2024eli}.}
\label{osci-parameter}
\end{table}
\begin{figure}[htbp!]
    \centering
    \subcaptionbox{\label{fig:sub1}}{\includegraphics[width=0.31\linewidth,height=5.8cm]{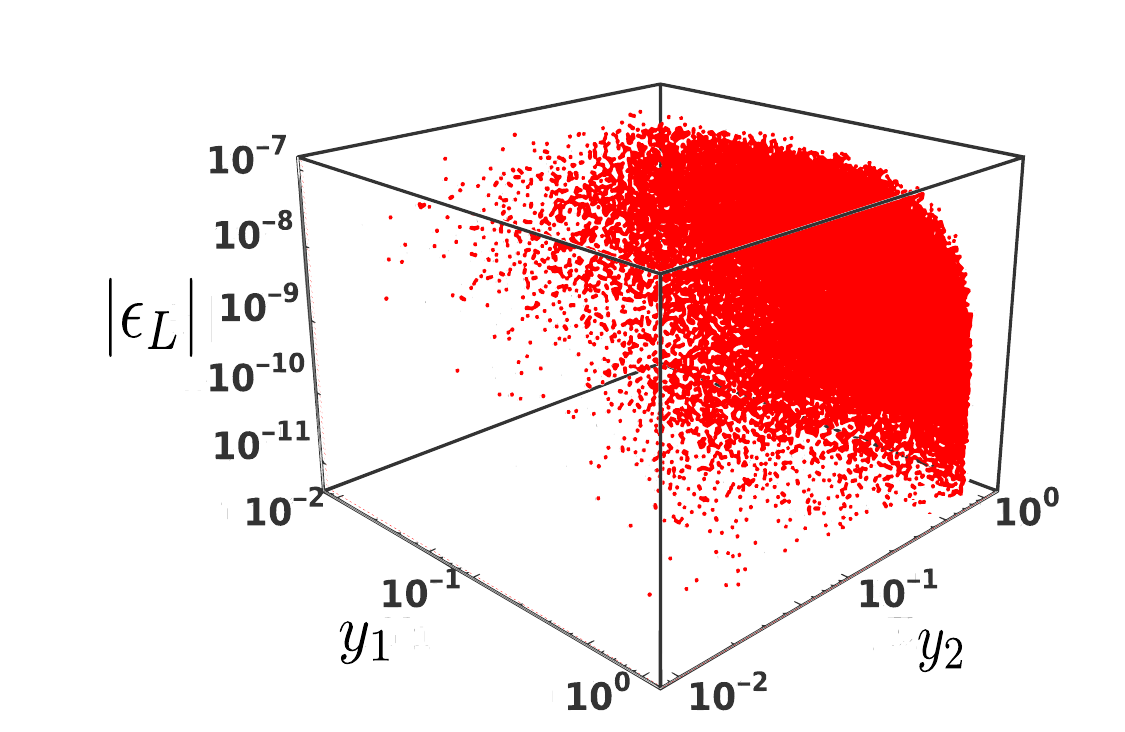}}
    \subcaptionbox{\label{fig:sub2}}{\includegraphics[width=0.31\linewidth,height=5.8cm]{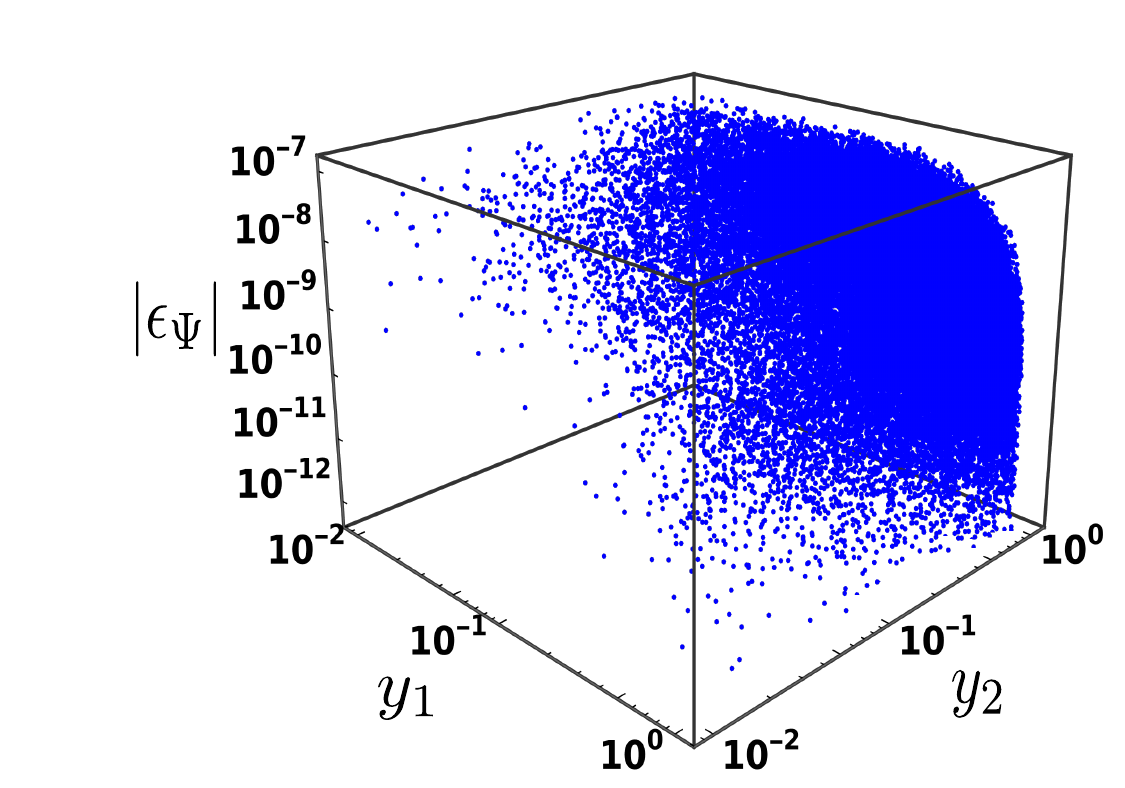}}
    \hspace{0.4cm}
    \subcaptionbox{\label{fig:sub3}}{\includegraphics[width=0.30\linewidth,height=5.8cm]{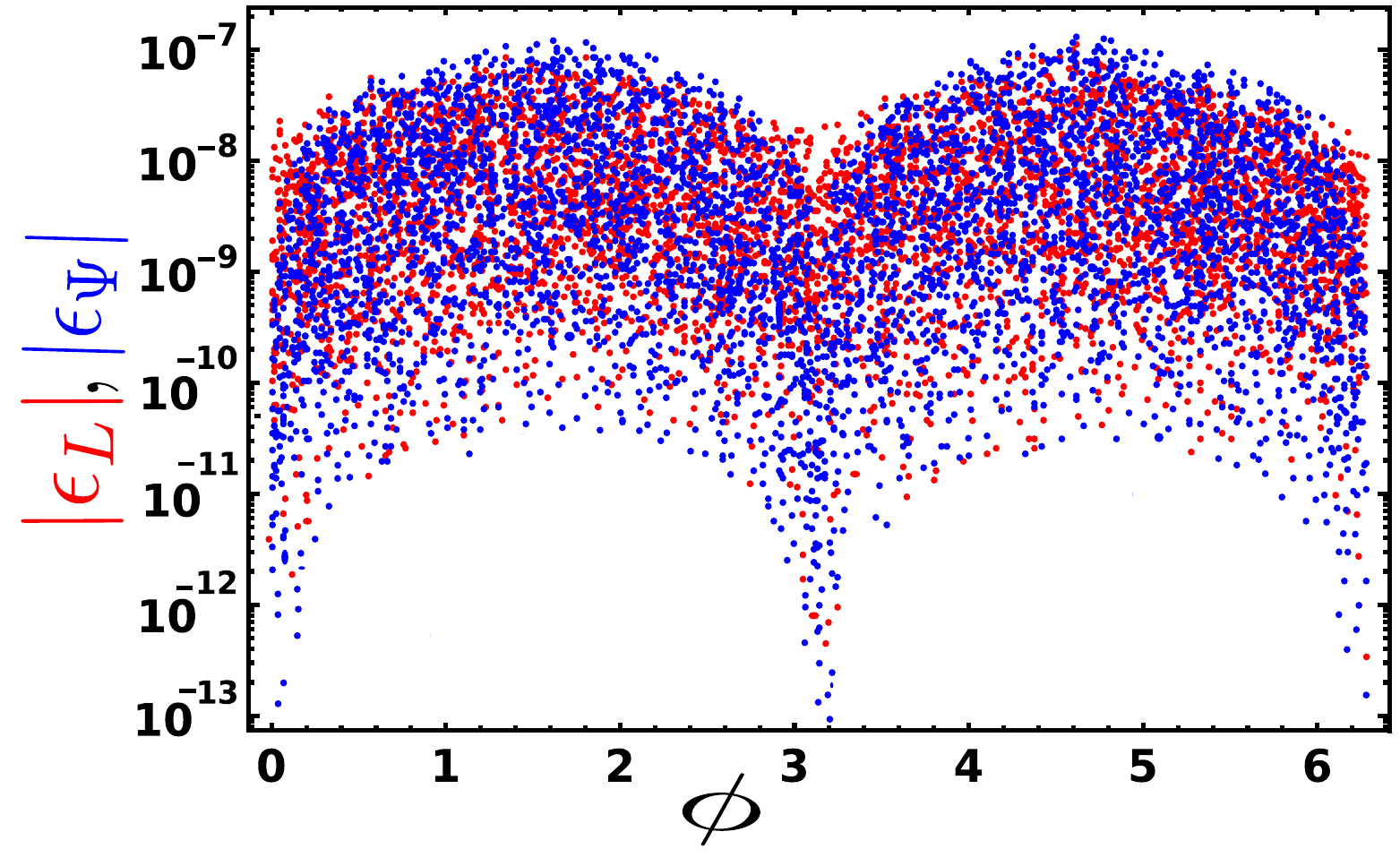}}
    \caption{The figures present the possible range of CP asymmetries $\epsilon_{L(\Psi)}$. The scattered plots are obtained by varying the expressions in equations (\ref{epsion-l})  and (\ref{epsilon-psi}) with respect to the Yukawa parameters $y_{1,2}$ between the range of the order $\sim(0.01$ - $1.0)$ (for figures (\ref{fig:sub1}) and (\ref{fig:sub2})), whereas the Fig.(\ref{fig:sub3}) is obtained with respect to the CP violating phase $\phi \in [0,2\pi]$. These plots are based on our model parameter space adhering to their specified values and the associated limiting conditions, which are set in accordance with the light neutrino mass constraint described in section (\ref{result}). The range for both the asymmetries is coming out to be $\epsilon_{L(\Psi)} \lesssim 10^{-7}$, and they exhibit a periodic behavior with respect to phase $\phi$.}
    \label{fig:set1}
\end{figure}
\begin{figure}[htbp!]
    \centering
    \subcaptionbox{\label{fig:sub4}}{\includegraphics[width=0.33\linewidth,height=5.2cm]{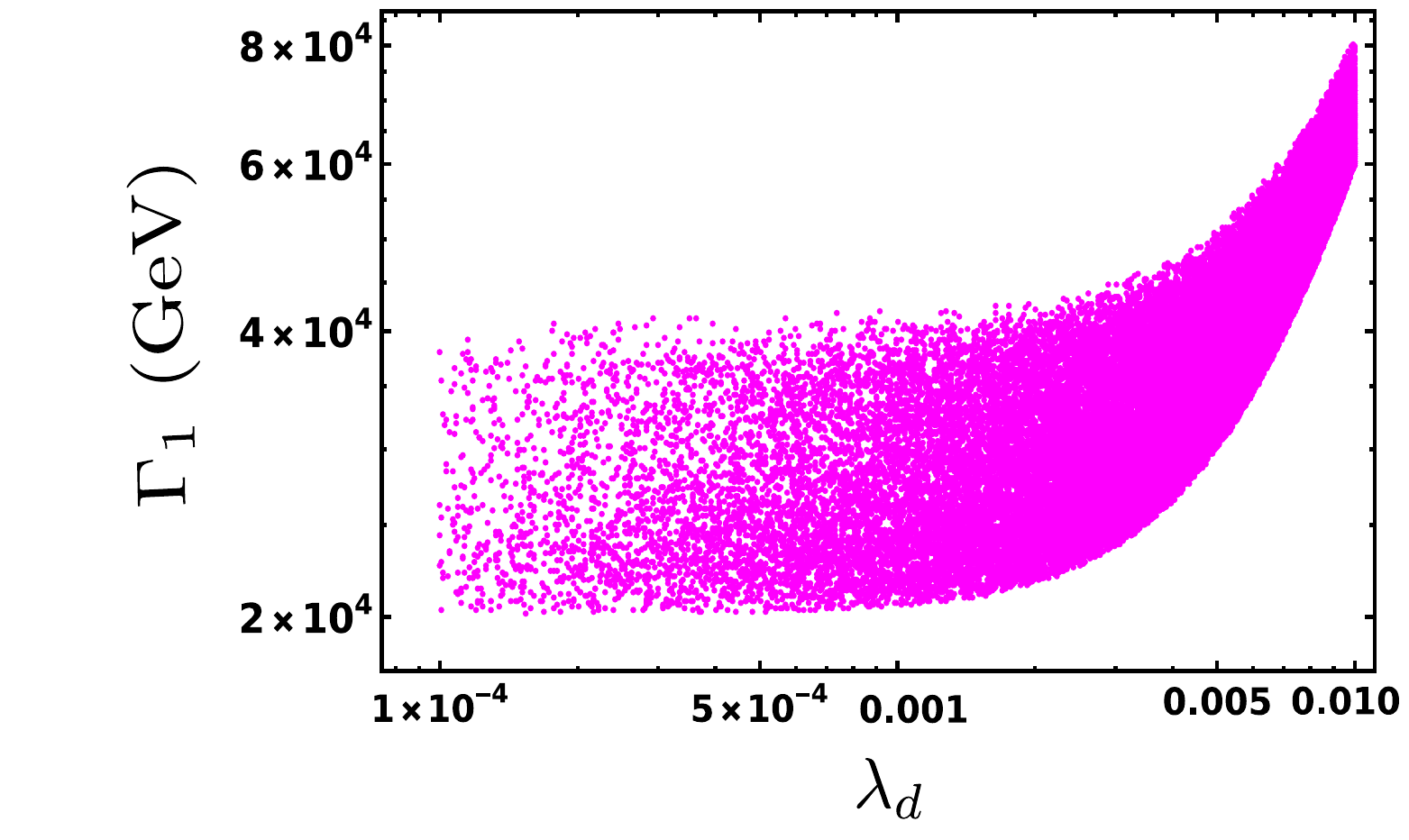}}
    \subcaptionbox{\label{fig:sub5}}{\includegraphics[width=0.31\linewidth,height=5.1cm]{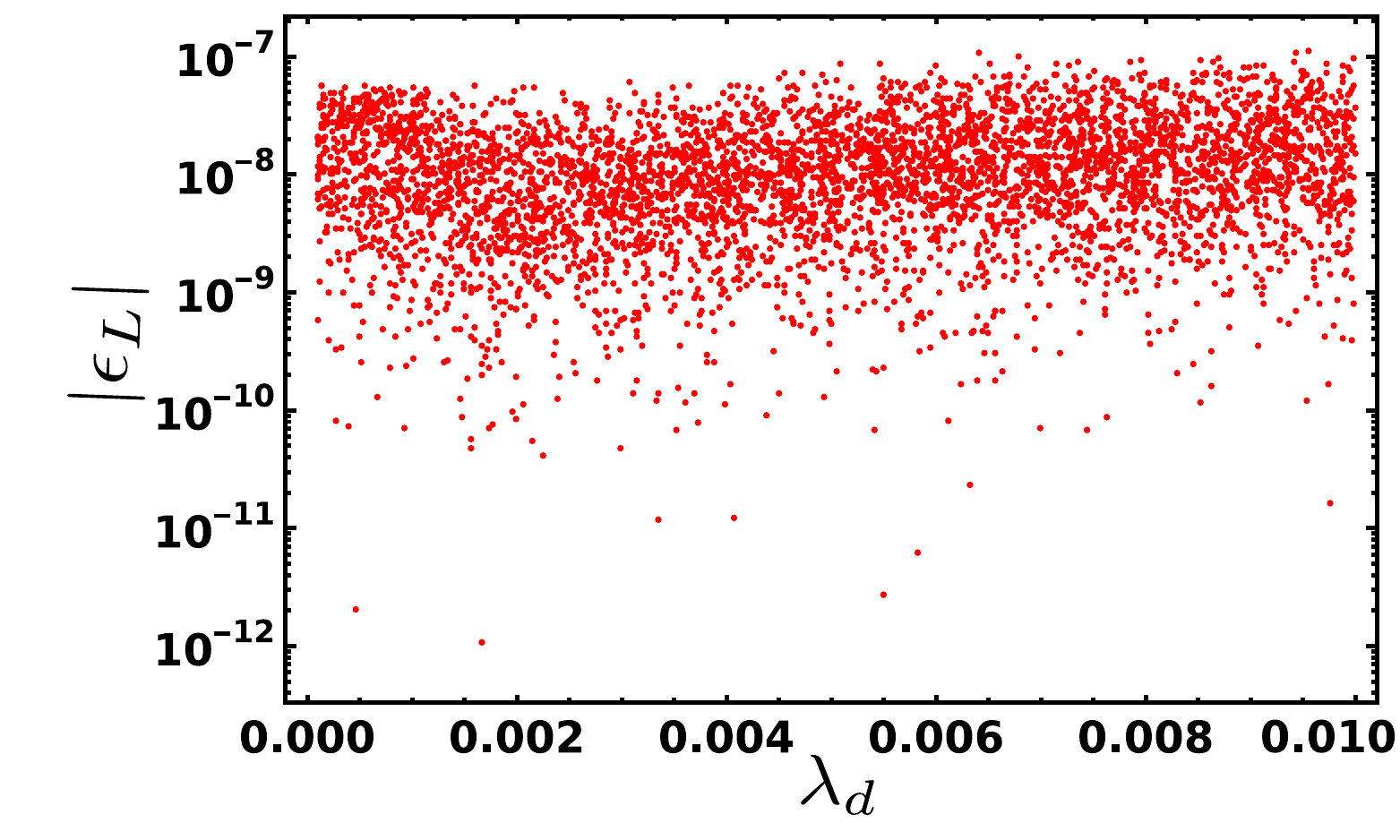}}
    \subcaptionbox{\label{fig:sub6}}{\includegraphics[width=0.32\linewidth,height=5.2cm]{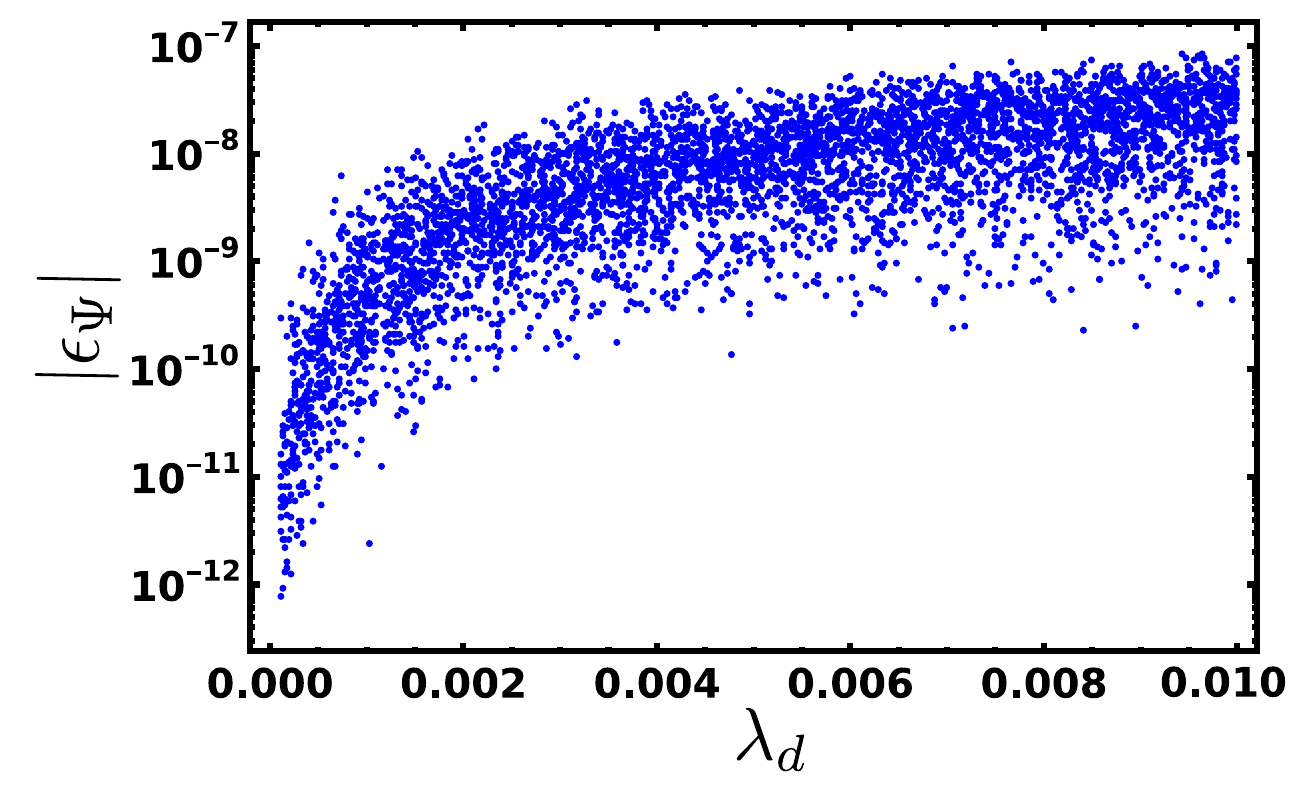}}
    \caption{The figures illustrate the variation of total decay width of lightest RHN $N_1$ i.e., $\Gamma_1$ and CP asymmetries $\epsilon_{L(\Psi)}$ with respect to the dark sector coupling $\lambda_d \in [10^{-4} - 10^{-2}]$ within our model parameter space and associated conditions. Fig.(\ref{fig:sub4}) gives the decay width $\Gamma_1 \sim 10^4$ GeV order, thus validating our assumption of narrow width approximation for our choice of RHN mass $M_1 \sim  10^{10}~\rm{GeV}$. The figures (\ref{fig:sub5}) and (\ref{fig:sub6}) illustrate the attainable range for the asymmetries in both the sectors with respect to $\lambda_d$.}
    \label{fig:set1.1}
\end{figure}

Now we proceed to define the variational range for our model parameters. We take the lightest right-handed neutrino mass required for successful leptogenesis as $M_1 \sim 10^{10}~\rm{GeV}$, which can be realized through the Eq.(\ref{MR1}), for $M \sim 10^{12}~\rm{GeV}, a \simeq 10^{-2}~ \rm{and}~ b \simeq 10^2$ with the hierarchical pattern, $M_1 < M_2 < M_3$. For $M_1 \sim 10^{10}~\rm{GeV}$, we take the Yukawa parameter $y^{\nu}_3 \sim 0.01$, the quartic coupling constant $\lambda^{\Phi \eta}_3 \sim 0.1$ \cite{Ahn:2013mva}. Subjected to the light neutrino mass constraint $\lesssim 10^{-11}~\rm{GeV}$, we can assign an individual variation of the order $\sim(0.01$ - $1.0)$ for the other Yukawa parameters $y_{1,2}$, by imposing the condition $ y^\nu_1 \neq y^\nu_2 \neq y^\nu_3$, required  for the off-diagonal elements of the combination $\tilde{Y}^{\dagger}_{\nu}\tilde{Y}_{\nu}$ to be non-zero (see Eq.(\ref{ynuynu})), which is crucial for leptogenesis. The values of the CP asymmetry parameters also depend on the phases $\psi_{1,2}$, which can be seen from the analytic expressions of $\epsilon_{L (\Psi)}$ presented in equations (\ref{analytic-epsilon-l}) and (\ref{analytic-epsilon-psi}). These phases are directly connected to the CP-violating phase $\phi$ and the parameter $\kappa$ (as shown in Eq.(\ref{alphs_beta})). The CP violating phase $\phi$ can have an independent variation from $0$ to $2\pi$ and for $\kappa=\lambda^s_\chi v_\chi/M$, we take the coupling constant $\lambda^s_{\chi} \sim10^{-4}$ with  $v_\chi \sim 10^{14} ~\rm{GeV}$, i.e., VEV of the $SU(2)_L$ singlet scalar $\chi$, complying with the light neutrino mass constraint. For the determination of loop function parameter $z_i$ (described in Sec.(\ref{sec:nu-matrix})), we assume the same mass of neutral component of IHD, i.e., $ \bar{m}_{\eta_i} \in  [80, 500]~\rm{GeV}$. In accordance with Eq.(\ref{eq-gamma1}), larger value of $\lambda_d$ will violate the narrow width approximation ($\Gamma_1 \ll M_1$) for our choice of lightest right-handed neutrino mass $M_1 \sim 10^{10}$~ GeV. Adhering to the narrow width approximation and simultaneously aiming for adequate dark sector asymmetry production, the value of the dark sector coupling is restricted in the range $\lambda_d \in [10^{-4} - 10^{-2}]$.

We have configured our model parameter space, its values, and the restricting conditions associated with them. Now, we use the mentioned conditions and perform a parameter scan to probe the other variables. By doing this, the scattered plots shown in Figs. (\ref{fig:set1}) and (\ref{fig:set1.1}) are obtained. The Fig.(\ref{fig:set1}) illustrates the all possible range for the lepton asymmetry and dark sector asymmetry, which are coming out to be of the order $\epsilon_{L(\Psi)} \lesssim 10^{-7}$. In the figures (\ref{fig:sub1}) and (\ref{fig:sub2}), the asymmetries are varied with respect to $y_{1,2}$ following the non-degeneracy of $y^\nu_1,~y^\nu_2$, and $y^\nu_3$. Fig.(\ref{fig:sub3}) depicts the variation of the asymmetries with respect to the CP-violating phase $\phi$. The red points correspond to the lepton asymmetry $\epsilon_L$ and the blue points correspond to the dark sector asymmetry $\epsilon_\Psi$. However, both the asymmetries exhibit periodic behavior with respect to $\phi$. At regular intervals, that means for $\phi = n\pi, n=0,1,2,\cdots$, we are getting minuscule asymmetries from both the sectors. This kind of periodic behavior intricately indicates the sub-dominance of the cosine terms containing phases $\psi_{1,2}$ from the analytic expressions shown in equations (\ref{analytic-epsilon-l}) and (\ref{analytic-epsilon-psi}).  In Fig.(\ref{fig:set1.1}), we tried to present the impact of dark sector coupling $\lambda_d$ on the variables. As per eq.(\ref{eq-gamma1}), the decay width $\Gamma_1$ is varied with respect to $\lambda_d \in [10^{-4} - 10^{-2}]$. As shown in Fig.(\ref{fig:sub4}), we are getting $\Gamma_1 \sim 10^{4}$~GeV, which validates the condition for narrow width approximation for our choice of RHN mass $M_1 \sim 10^{10}~\rm{GeV}$. The figures (\ref{fig:sub5}) and (\ref{fig:sub6}) illustrate the attainable range for the asymmetries in the lepton and dark sector, respectively, with respect to 
$\lambda_d$. Notably, the dark sector asymmetries are gradually evolving, indicating feeble washout effects as compared to the lepton sector.  
\begin{figure}[htbp!]
    \centering
    \includegraphics[width=6.5cm,height=5.2cm]{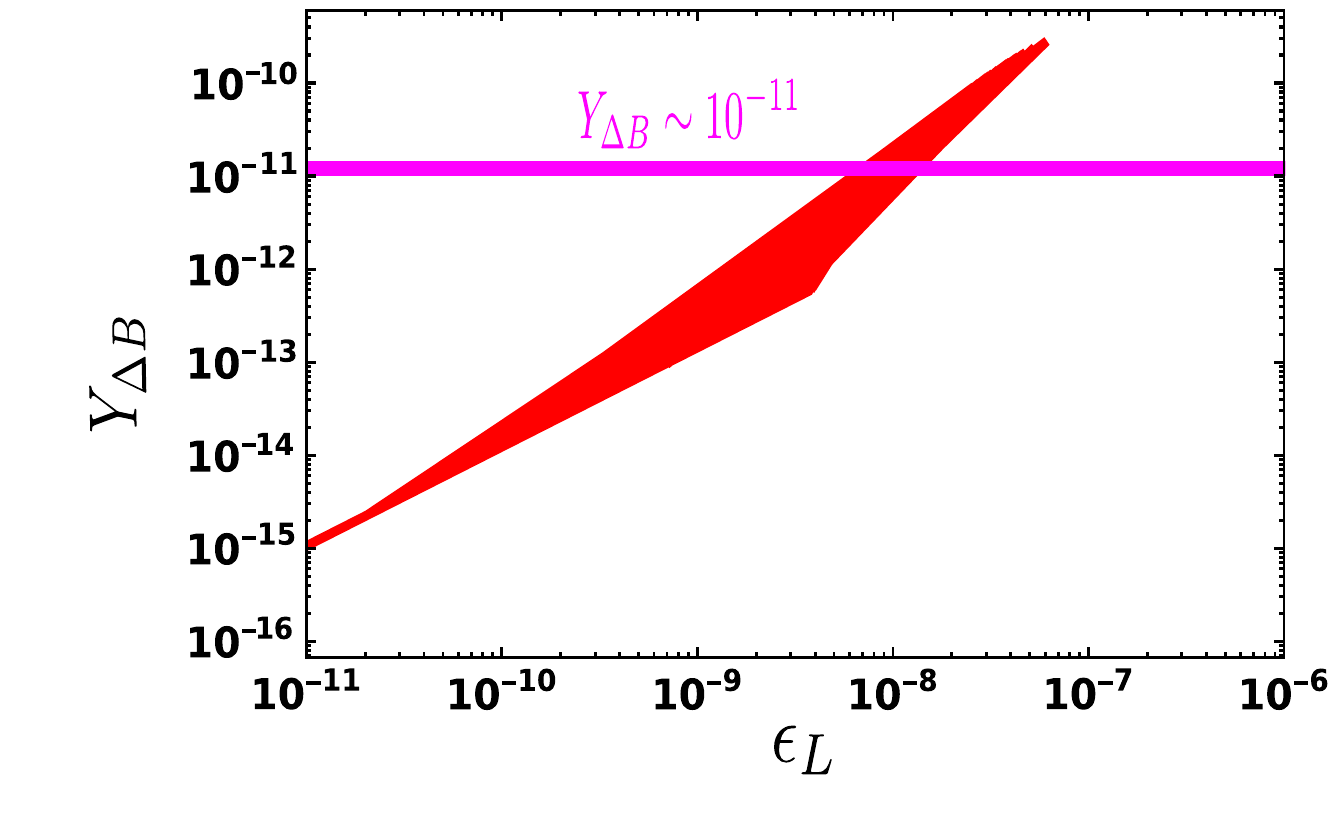}
    \caption{Figure shows the parallel mapping of the abundance yield $Y_{\Delta B}$ for the visible sector corresponding to its CP asymmetry values $\epsilon_L$. The magenta line shows the observable final abundance yield for baryons at present temperature, which is of the order $\sim 10^{-11}$.}
    \label{fig:set1.2}
\end{figure}
\begin{figure}
    \centering
    \subcaptionbox{\label{fig:sub8.1}}{\includegraphics[width=4.9cm,height=5.2cm]{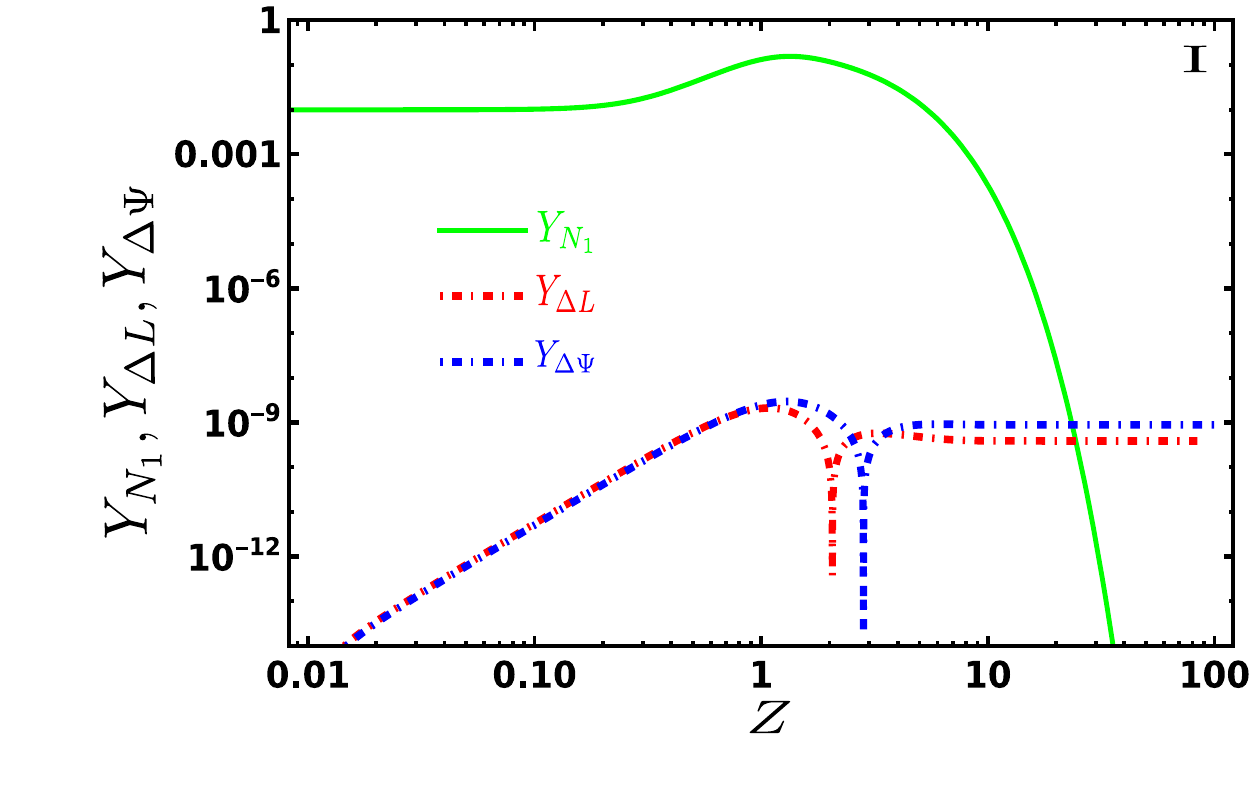}}
    \hspace{0.1cm}
    \subcaptionbox{\label{fig:sub8.2}}{\includegraphics[width=4.9cm,height=5.0cm]{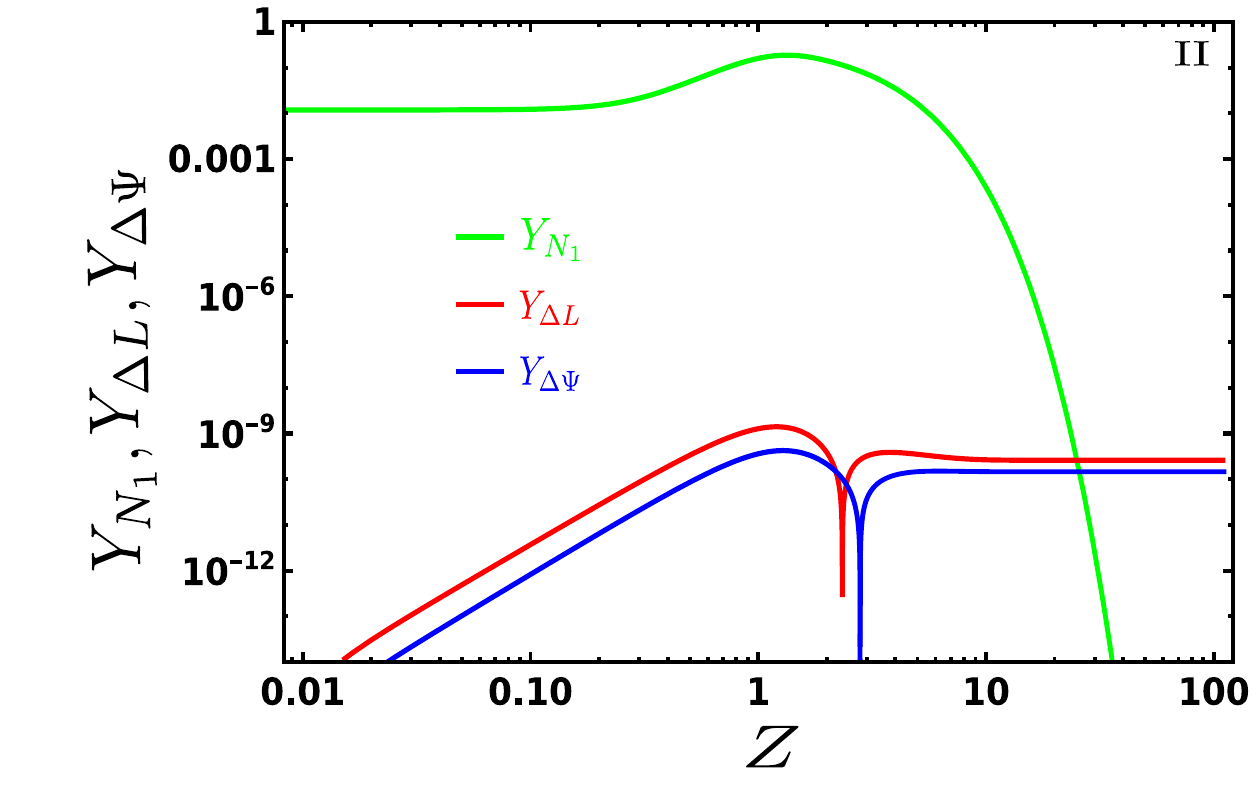}}
    \hspace{0.1cm}
    \subcaptionbox{\label{fig:sub8.3}}{\includegraphics[width=4.9cm,height=5.2cm]{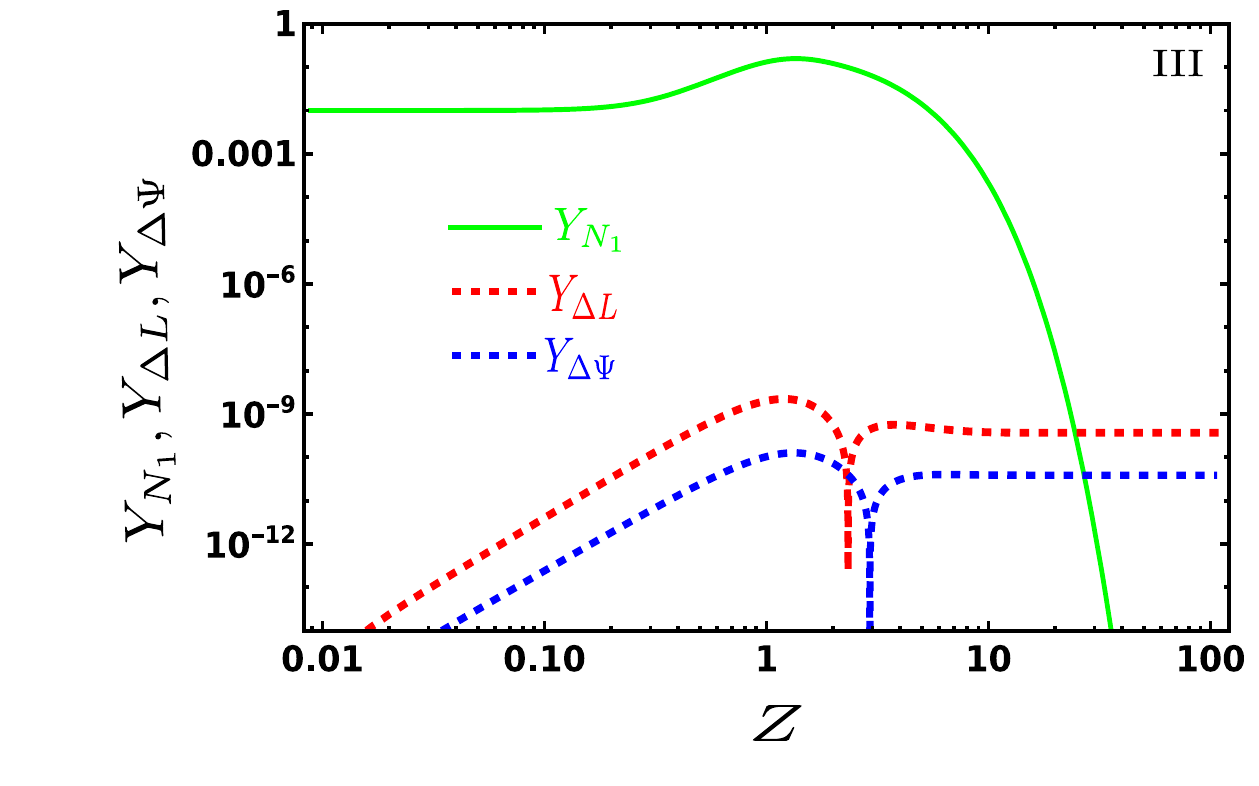}}
    \caption{The figures illustrate the solutions of the Boltzmann equations. The green curve presents the evolution of $Y_{N1}$ with respect to $z=M_1/T$ by using Eq.(\ref{eq:BE}). The red curve and blue curve represent the evolution of asymmetry abundances in the visible(dark) sector $Y_{\Delta L(\Psi)}$, respectively, by using Eq.(\ref{eq:BE1}). Different curve styles are used to distinguish the chosen combinations of $\epsilon$ values (dotdashed, solid, and dashed lines correspond to combinations I, II, and III in the same order). The chosen combinations of CP asymmetry values $(\epsilon_L, \epsilon_\Psi)$ are: $\rm{I}-(2\times10^{-8},2\times10^{-8}),~\rm{II}-(1.25\times10^{-8},3.38\times10^{-9})$, and $\rm{III}-(1.12\times10^{-8},1.03\times10^{-9})$.}
    \label{fig:set1.3}
\end{figure}
\begin{figure}
    \centering
    \includegraphics[width=9cm,height=5.2cm]{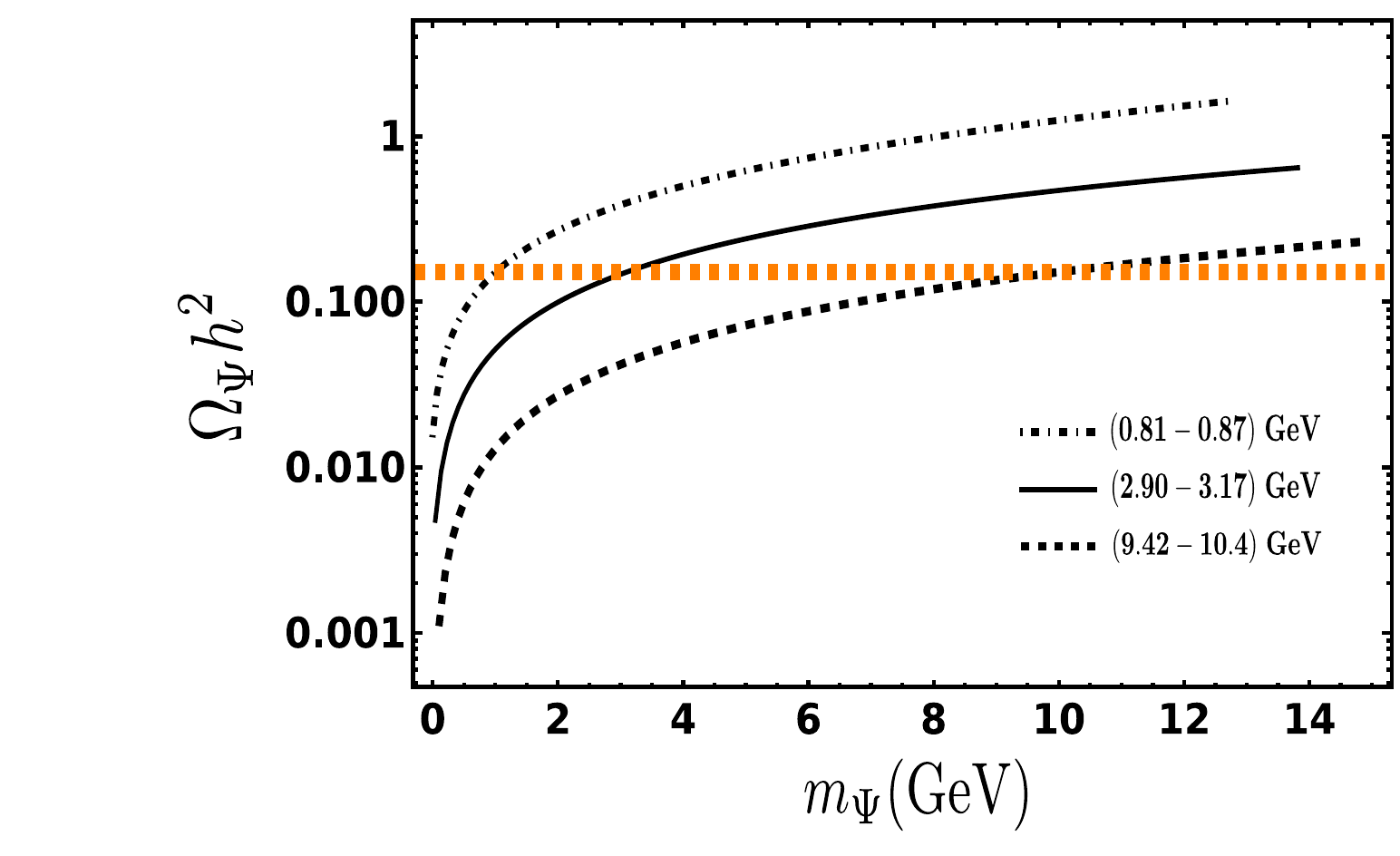}
    \caption{ Figure shows the attainable DM relic density for different mass ranges of dark fermion $\Psi$. Using the value of the dark sector abundance yield $Y_{\Delta \Psi}$ at present temperature, in Eq.(\ref{dark-relic}), variation of $\Omega_{\Psi}h^2$ with respect to $m_\Psi$ is depicted. Different curve styles are used to distinguish the chosen combinations of $\epsilon$ values (dotdashed, solid, and dashed lines correspond to combinations I, II, and III in the same order). The chosen combinations of CP asymmetry values $(\epsilon_L, \epsilon_\Psi)$ are: $\rm{I}-(2\times10^{-8},2\times10^{-8}),~\rm{II}-(1.25\times10^{-8},3.38\times10^{-9})$, and $\rm{III}-(1.12\times10^{-8},1.03\times10^{-9})~$. For the mentioned mass range corresponding to different combinations, the  $\Omega_{\Psi}h^2$ value matches the experiment limit $ \Omega_{DM}h^2 = 0.12 \pm 0.001$, satisfying $\Omega_{DM}/\Omega_{b} \sim 5$.}
    \label{fig:set1.4}
\end{figure}

As of yet, we have examined the amount of CP asymmetry generated in both sectors and obtained the order of decay width of $N_1$. Moving forward, we now investigate the evolution of asymmetry abundance using the Boltzmann equations (\ref{eq:BE}) and (\ref{eq:BE1}). For solving the Boltzmann equations, we can use $Y^{\rm{eq}}_{N_1} = n^{\rm{eq}}/s$ with $n^{\rm{eq}} = \frac{g}{2\pi^{2}} M_1^{2} T  K_{2}(z)$ \cite{PhysRevD.45.455}. 
To produce more DM density than BAU, we work with a weaker washout effect in the dark sector than the visible sector (according to Fig.(\ref{fig:sub6})), we can consider $Br_\Psi < Br_L$). Then we have performed a parallel mapping in the plane $Y_{\Delta B} - \epsilon_L$ by solving the Boltzmann equation corresponding to values of $\epsilon_L$ obtained within our model parameter space. This mapping is shown in the Fig.(\ref{fig:set1.2}), which infers the disfavored range of $ \epsilon_L \gtrsim 10^{-8}$ for which the observable present day baryon asymmetry of the universe ($Y_{\Delta B}$) exceeds the order of $10^{-11}$. These facts inform our choice to consider $\epsilon_L \sim 10^{-8}$. For subsequent calculations, we will consider different 
$\epsilon_\Psi$ value keeping $\epsilon_L \sim 10^{-8}, \Gamma_1 \sim 10^{4} ~\rm{GeV}$ and $M_1 \sim 10^{10}$~ GeV. From our generated data set, we have chosen three combinations of CP asymmetry values $(\epsilon_L, \epsilon_\Psi)$: $\rm{I}-(2\times10^{-8}, 2\times10^{-8}), \rm{II}-(1.25\times10^{-8}, 3.38\times10^{-9})$,
and $\rm{III}-(1.12\times10^{-8}, 1.03\times10^{-9})$. The choice of combination I is based on the premise that we are getting $\epsilon_{L(\Psi)} \lesssim 10^{-7}$. So it will be reasonable to investigate the impact of the same order of CP asymmetry on the present-day baryon asymmetry. The rest two combinations are based on the fact that the washout effects in the dark sector are weaker than the visible sector within our model parameter space, and hence the combination ($\epsilon_L \sim 10^{-8}, \epsilon_\Psi \sim 10^{-9}$) appears to be a justified combination to be investigated. 

\begin{table}[h!]
\centering
\renewcommand{\arraystretch}{1.4} 
\begin{tabularx}{\textwidth}{|c|Y|c|c|c|c|}
\hline
Sets & \boldmath$(\epsilon_L, \epsilon_\Psi)$ & \boldmath$Y_{\Delta L}$ & \boldmath$Y_{\Delta B}$ & \boldmath$Y_{\Delta \Psi}$ & \boldmath$m_\Psi~\rm{(GeV)}$ \\
\hline
I & $(2\times10^{-8},~2\times10^{-8})$ & $3.85 \times 10^{-10}$ & $1.34 \times 10^{-10}$ & $7.59 \times 10^{-10}$ & $0.81 - 0.87$ \\
II & $(1.25\times10^{-8},~3.38\times10^{-9})$ & $2.40 \times 10^{-10}$ & $8.38 \times 10^{-11}$ & $1.28 \times 10^{-10}$ & $2.90 - 3.17$ \\
III & $(1.12\times10^{-8},~1.03\times10^{-9})$ & $2.15 \times 10^{-10}$ & $7.51 \times 10^{-11}$ & $3.87 \times 10^{-11}$ & $9.42 - 10.4$ \\
\hline
\end{tabularx}
\caption{The table shows the values of present-day final abundance yields $Y_{\Delta L(\Psi)}$ for the visible (dark) sector and the corresponding baryon asymmetry of the universe $Y_{\Delta B}$ for the three combinations of $(\epsilon_L,\epsilon_\Psi)$. The mass of dark fermion $\Psi$ for which the DM relic density $\Omega_\Psi h^2$ matches its experiment limit $\Omega_{DM}h^2 = 0.12 \pm 0.001$ and satisfying $\Omega_{DM}/\Omega_{b} \sim 5$ is also presented.}
\label{tab:results}
\end{table}

With the above chosen combination of CP asymmetry values, we solved the Boltzmann equations, and the resulting abundance yields are depicted in Fig.{\ref{fig:set1.3}}, where the green curve represents the evolution of $Y_{N1}$ with respect to $z=M_1/T$ by using Eq.(\ref{eq:BE}). The red curve and blue curve represent, correspondingly, the evolution of abundance yield in the visible sector $Y_{\Delta L}$ and dark sector $Y_{\Delta \Psi}$ by using Eq.(\ref{eq:BE1}). Different curve styles are used to distinguish the chosen combinations of asymmetries (dotdashed, solid, and dashed lines correspond to combinations I, II, and III, respectively). 
For the respective combinations of $\epsilon_{L(\psi)}$, the obtained values of the final abundance yields $Y_{\Delta L(\Psi)}$ in lepton(dark) sector at present day temperature and the corresponding baryon asymmetry value $Y_{\Delta B}$ are given in the Table-(\ref{tab:results}).
Although we have considered $Br_\Psi < Br_L$ for all the combinations, as can be seen from Fig.(\ref{fig:sub8.1}), we are getting more abundance yields for the dark sector than the visible sector for equal order of CP asymmetries. For the rest of the combinations, $Y_{\Delta \Psi} < Y_{\Delta L}$ is achieved as per the expectation (shown in the Figs.(\ref{fig:sub8.2}) and (\ref{fig:sub8.3})). After obtaining the $Y_{\Delta L}$ from each combination, we have calculated the baryon asymmetry $Y_{\Delta B}$ at present temperature with the help of Sphaleron conversion shown in eq.(\ref{sphaleron}). 
As inferred from the Table-(\ref{tab:results}), the $Y_{\Delta B}$ value for combination-I exceeds the experimentally observed amount by one order of magnitude, whereas for the combination-II, the $Y_{\Delta B}$ value comply well with the experimentally observed value than that for the combination-III. Now using the obtained $Y_{\Delta \Psi}$ values in Eq. (\ref{dark-relic}), we will analyse the dark matter relic density and the corresponding mass of the dark matter candidate in our model. Fig.(\ref{fig:set1.4}) shows the amount of dark matter relic density obtained for the mass range $m_\Psi \in [0.1 - 14]$~GeV within our model parameter space. We have found three specific mass range corresponding to the three combinations ((0.81 - 0.87)~GeV,~(2.90 - 3.17)~GeV,~(9.42 - 10.4)~GeV for combination I, II, III respectively) for which our result of $\Omega_{\Psi}h^2$ is matching the observed value $\Omega_{DM}h^2 = 0.12 \pm 0.001$ and satisfying $\Omega_{DM}/\Omega_{b} \sim 5$.

The direct detection aspect of the DM candidate $\Psi$ can be understood from its coupling to 
the visible sector. The asymmetric DM field does not couple to the SM Higgs and $Z$-bosons. It can 
do so via an effective vertex, $h\bar{\Psi}\Psi$ at one loop, as a result, the leading order scattering cross-section
is highly suppressed. Hence, the asymmetric DM particle escapes all the direct detection bounds.
\section{Conclusion}
\label{conclusion}
The neutral component of the IHD in a two-Higgs doublet model acts as a potential dark matter candidate.
It plays a subdominant role in matching the observed relic density for a specific mass range between $(80 - 500)$ GeV, hence making this our region of interest to probe whether the deficit can be compensated by other DM candidates. In this regard, we have proposed a renormalizable model based on the gauge group $SU(2)_L \times U(1)_Y \times A_4 \times Z_2 \times Z^{\prime}_2 \times Z_3$ to 
provide a unified framework for producing the light neutrino mass,
relic density of visible and dark matter in similar order of magnitude.
Here, we work in the framework of two-sector leptogenesis.
The model suitably accounts for spontaneous CP violation by the complex VEV of a scalar field. The single CP-violating phase associated with the VEV connects CP violation in the low- and high-energy sectors, hence reducing the number of free parameters in the Lagrangian. 
The particle content of our model is incorporated in two sectors: visible and dark sectors. 
The visible sector contains one inert Higgs doublet ($\eta$), two singlet scalars ($\chi,~\chi^\prime$), three RH neutrinos $N_{R_i}$ along with the SM Higgs doublet $\Phi$ and other fermions of the SM. Whereas, the dark sector comprises a fermion singlet $\Psi$ and a real scalar singlet $S$. A flavor symmetric realization, accounting for a predictive flavor structure and spontaneous CP violation, is achieved through the $A_4$ discrete symmetry group. 
The auxiliary symmetry groups $Z_2, Z^{\prime}_2, Z_3$~ are there to ensure the stabilization of the extended particle content. 
These discrete symmetries collectively account for the light neutrino mass generation by through the radiative seesaw mechanism and circumvent the vacuum alignment problem. 
In our model, the spontaneous CP violation is accounted for by the phase $\phi$ associated with the VEV of $\chi$. 

In this context, we wanted to replenish the dark matter population by the decay of RHNs in a two-sector leptogenesis scenario. This enables us to figure out the CP asymmetries in both the sectors from the decay of lightest RH neutrino $N_1$ into SM lepton doublets and IHD in the visible sector and into the fermion $\Psi$ and scalar $S$ in the dark sector simultaneously. Subjected to the connection between $N_1$ and the two sectors through Yukawa coupling only, we have adopted a particular type of Yukawa construction $\tilde{Y}_\nu$, emerging from $A_4$ symmetry breaking, having complex entries which acts as the sole contributor to the source of CP violation and kept the Yukawa coupling with the dark sector real for the sake of simplicity. We then configured our model parameters and their operational range in accordance with the light neutrino mass constraint (obtained using oscillation data for NH here) and the non-zero off-diagonal elements for the combination  $\tilde{Y}^{\dagger}_{\nu}\tilde{Y}_{\nu}$.

In a numerical analysis, using our model parameter space, we have obtained the CP asymmetry values for both the sectors of the order $\epsilon_{L(\Psi)} \lesssim 10^{-7}$ for $M_1 \sim 10^{10}~\rm{GeV}$. The variation of $\epsilon_{L(\Psi)}$ with respect to the CP violating phase $\phi$ shows the periodic behavior at $\phi = n\pi, n=0,1,2,\cdots$ emphasizing the sub-dominance of the terms with cosine factor and large mass hierarchy of RHNs, hence ensures that both sectors have common dependence on the phase $\phi$. We have also observed the weaker washout effects in the dark sector than the visible sector with respect to the dark sector coupling $\lambda_d$ in the limit of narrow width approximation (as inferred from the obtained value of $\Gamma_1 \sim 10^4$ GeV corresponding to $M_1 \sim 10^{10}$ GeV). Then we have solved the Boltzmann equations for different combinations of $\epsilon$ values assuming $Br_\Psi < Br_L$ for all and studied the evolution of abundance yield in both the sectors. This provides us the final abundance yields $Y_{\Delta L(\Delta \Psi)}$ at present day temperature, using which we have calculated the present day baryon asymmetry $Y_{\Delta B}$ and also explored the mass range $m_{\psi}$ of dark fermion for which non-zero DM relic density can be achieved. Among the three combinations considered, the combination with same order of $\epsilon_{L(\Psi)} \sim 10^{-8}$ value shows gradual enhancement in the $Y_{\Delta \Psi}$ value towards present day temperature and the corresponding relic density coincide with the experimental limit in a mass range of (0.81 - 0.87) GeV satisfying $\Omega_{DM}/\Omega_{b} \sim 5$. For the remaining two combinations, where we have considered the $\epsilon$ values with one order of magnitude difference ($\epsilon_L \sim 10^{-8}, \epsilon_\Psi \sim 10^{-9}~$), the evolution of final yields in both the sector follows $Y_{\Delta L} > Y_{\Delta \Psi}$ at present day temperature and the corresponding mass range of the dark fermion, for which $\Omega_{DM}/\Omega_{b} \sim 5$ is validated, is bit larger than that for the first combination. Overall, we can say that our model parameter space accounts for the observed DM relic density for a few GeV order of mass of the dark sector fermion. 
\appendix
\section{Multiplication rule in $A_4$ symmetry}
\label{app:mult-rule}
$A_4$ is the finite group of even permutations of four objects and is the smallest non-abelian group with a three-dimensional irreducible representation \cite{Ma:2001dn, Ma:2004zv}. It has 12 elements and four irreducible representations: one triplet ${\bf 3}$ and three singlets (${\bf 1}, {\bf 1}', {\bf 1}''$). The multiplication rule followed by them is given by
\begin{eqnarray}
{\bf 1}' \otimes {\bf 1}'' &=& {\bf 1} \,, \nonumber \\
{\bf 1}' \otimes {\bf 1}' &=& {\bf 1}'' \,, \nonumber \\
{\bf 1}'' \otimes {\bf 1}'' &=& {\bf 1}' \,  \nonumber \\
{\bf 3} \otimes {\bf 3} &=& {\bf 3}_s \oplus {\bf 3}_a \oplus {\bf 1} \oplus {\bf 1}' \oplus {\bf 1}'' \,.
\end{eqnarray}
If a and b be two $A_4$ triplets such that $a=(a_{1}, a_{2}, a_{3})$ and $b=(b_{1}, b_{2}, b_{3})$, then the decomposition will be
\begin{eqnarray}
  (a\otimes b)_{{\bf 3}_{\rm s}} &=& (a_{2}b_{3}+a_{3}b_{2}, a_{3}b_{1}+a_{1}b_{3}, a_{1}b_{2}+a_{2}b_{1})~,\nonumber\\
  (a\otimes b)_{{\bf 3}_{\rm a}} &=& (a_{2}b_{3}-a_{3}b_{2}, a_{3}b_{1}-a_{1}b_{3}, a_{1}b_{2}-a_{2}b_{1})~,\nonumber\\
  (a\otimes b)_{{\bf 1}} &=& a_{1}b_{1}+a_{2}b_{2}+a_{3}b_{3}~,\nonumber\\
  (a\otimes b)_{{\bf 1}'} &=& a_{1}b_{1}+\omega a_{2}b_{2}+\omega^{2}a_{3}b_{3}~,\nonumber\\
  (a\otimes b)_{{\bf 1}''} &=& a_{1}b_{1}+\omega^{2} a_{2}b_{2}+\omega a_{3}b_{3}~,
 \end{eqnarray}
where the complex number $\omega$ is the cube root of unity i.e., $\omega=e^{i2\pi/3}$.

\section{Scalar potential under $  SU(2)_L \times U(1)_Y \times A_4 \times Z_2 \times Z_2'\times Z_3$}
\label{app:potential}
The scalar potential under considered gauge symmetry $  SU(2)_L \times U(1)_Y \times A_4 \times Z_2 \times Z_2' \times Z_3$ is given by,
\begin{equation}
    V(\Phi)= \mu_{\Phi}^{2} \left( \Phi^{\dagger} \Phi \right) + \lambda_{\Phi} (\Phi^{\dagger} \Phi)^2 .
\end{equation}
\begin{equation}
    V(\eta) = \mu_{\eta}^{2} \left( \eta^{\dagger} \eta \right) + \lambda_{\eta} (\eta^{\dagger} \eta)^2.
\end{equation}
\begin{equation}
    V(S) = \mu_{S}^{2} \, (S^*S) + \lambda_{S} (S^*S)^4.
\end{equation}
\begin{eqnarray}
V(\chi) &=& \mu^{2}_{\chi}\left\{(\chi\chi)_{\mathbf{1}}+(\chi^{\ast}\chi^{\ast})_{\mathbf{1}}\right\}+m^{2}_{\chi}(\chi\chi^{\ast})_{\mathbf{1}}+\lambda^{\chi}_{1}\left\{(\chi\chi)_{\mathbf{1}}(\chi\chi)_{\mathbf{1}}+(\chi^{\ast}\chi^{\ast})_{\mathbf{1}}(\chi^{\ast}\chi^{\ast})_{\mathbf{1}}\right\} \nonumber\\
&+&\lambda^{\chi}_{2}\left\{(\chi\chi)_{\mathbf{1}^\prime}(\chi\chi)_{\mathbf{1}^{\prime\prime}}+(\chi^{\ast}\chi^{\ast})_{\mathbf{1}^\prime}(\chi^{\ast}\chi^{\ast})_{\mathbf{1}^{\prime\prime}}\right\} \nonumber\\
&+&\tilde{\lambda}^{\chi}_{2} \left\{(\chi^{\ast}\chi)_{\mathbf{1}^\prime}(\chi\chi)_{\mathbf{1}^{\prime\prime}}+(\chi^{\ast}\chi)_{\mathbf{1}^{\prime\prime}}(\chi^{\ast}\chi^{\ast})_{\mathbf{1}^{\prime}}\right\} \nonumber\\
  &+&\lambda^{\chi}_{3}\left\{(\chi\chi)_{\mathbf{3}_{s}}(\chi\chi)_{\mathbf{3}_{s}}+(\chi^{\ast}\chi^{\ast})_{\mathbf{3}_{s}}(\chi^{\ast}\chi^{\ast})_{\mathbf{3}_{s}}\right\}+\tilde{\lambda}^{\chi}_{3}(\chi^{\ast}\chi)_{\mathbf{3}_{s}}\left\{(\chi\chi)_{\mathbf{3}_{s}}+(\chi^{\ast}\chi^{\ast})_{\mathbf{3}_{s}}\right\}\nonumber\\
  &+&\lambda^{\chi}_{4}\left\{(\chi^\ast\chi)_{\mathbf{3}_{a}}(\chi\chi)_{\mathbf{3}_{s}}+(\chi\chi^{\ast})_{\mathbf{3}_{a}}(\chi^{\ast}\chi^{\ast})_{\mathbf{3}_{s}}\right\}\nonumber\\
  &+&\xi^{\chi}_{1}\left\{\chi(\chi\chi)_{\mathbf{3}_{s}}+\chi^{\ast}(\chi^{\ast}\chi^{\ast})_{\mathbf{3}_{s}}\right\}+\tilde{\xi}^{\chi}_{1}\left\{\chi(\chi^{\ast}\chi^{\ast})_{\mathbf{3}_{s}}+\chi^{\ast}(\chi\chi)_{\mathbf{3}_{s}}\right\}~.
 \end{eqnarray}
 \begin{align}
V(\chi') &=  m_{\chi'}^2 (\chi' \chi'^{\ast})_{\mathbf{1}} \nonumber \\
&\quad + \lambda_5^{\chi'} \left\{ (\chi'^{\ast} \chi')_{\mathbf{3}s} \cdot (\chi'^{\ast} \chi')_{\mathbf{3}s} \right\}_{\mathbf{1}} \nonumber\\
&\quad + \xi_1^{\chi'} \left\{ \chi' \cdot (\chi' \chi')_{\mathbf{3}s} + \chi'^{\ast} \cdot (\chi'^{\ast} \chi'^{\ast})_{\mathbf{3}s} \right\}.
\end{align}
\begin{align}
V(\Phi, \chi) &= \lambda_{\Phi\chi}^{(1)} (\Phi^{\dagger} \Phi)(\chi^{\ast} \chi)_{\mathbf{1}}
+ \lambda_{\Phi\chi}^{(2)} (\Phi^{\dagger} \Phi) \left[ (\chi \chi)_{\mathbf{1}} + (\chi^{\ast} \chi^{\ast})_{\mathbf{1}} \right].
\end{align}
\begin{align}
V(\Phi, \chi') &= \lambda_{\Phi\chi'}^{(1)} (\Phi^{\dagger} \Phi)(\chi'^{\ast}  \chi')_{\mathbf{1}}.
\end{align}
\begin{equation}
V(\Phi,\eta) = \lambda^{\Phi\eta}_{1} (\Phi^{\dagger} \Phi)(\eta^{\dagger} \eta) +\lambda^{\Phi\eta}_{2} (\phi^{\dagger} \eta)(\eta^{\dagger} \Phi)+ \lambda^{\Phi\eta}_{3}\left\{(\eta^{\dag}\Phi)(\eta^{\dag}\Phi)+h.c\right\}.
\end{equation}
\begin{equation}
    V(\Phi,S) = \lambda_{\Phi S} (\Phi^{\dagger} \Phi)(S^* S).
\end{equation}
\begin{align}
V(\eta, \chi) &= \lambda_{\eta\chi}^{(1)} (\eta^{\dagger} \eta)(\chi^{\ast} \chi)_{\mathbf{1}}
+ \lambda_{\eta\chi}^{(2)} (\eta^{\dagger} \eta) \left[ (\chi \chi)_{\mathbf{1}} + (\chi^{\ast} \chi^{\ast})_{\mathbf{1}} \right].
\end{align}
\begin{align}
V(\eta, \chi') &= \lambda_{\eta\chi'}^{(1)} (\eta^{\dagger} \eta)(\chi'^{\ast} \chi')_{\mathbf{1}}.
\end{align}
\begin{equation}
    V(\eta,S) = \lambda_{\eta S} (\eta^{\dagger} \eta) (S^* S).
\end{equation}
\begin{align}
V(\chi,\chi') &= \lambda^{\chi\chi'}_1 \, (\chi^{\ast} \chi)_{\mathbf{1}} \, (\chi'^{\ast} \chi')_{\mathbf{1}}
+ \lambda^{\chi\chi'}_2 \left[ (\chi^{\ast} \chi')_{\mathbf{3}s} \cdot (\chi'^{\ast} \chi)_{\mathbf{3}s} \right] \nonumber\\
&\quad + \lambda^{\chi\chi'}_3 \left[ (\chi^{\ast} \chi')_{\mathbf{1}'} (\chi'^{\ast} \chi)_{\mathbf{1}''} + \textrm{h.c.} \right].
\end{align}
\begin{equation}
    V(\chi, S) = \lambda_{\chi S}(\chi^{\ast} \chi)_{\mathbf{1}}(S^* S).
\end{equation}
\begin{equation}
    V(\chi', S) = \lambda_{\chi' S}(\chi'^{\ast} \chi')_{\mathbf{1}}(S^* S).
\end{equation}
Here $\mu_{\Phi},\mu_{\eta}, \mu_S, \mu_{\chi}, m_{\chi}, m_{\chi'}, \xi^{\chi}_1, \tilde{\xi}^{\chi}_1 ~\rm{and}~ \xi^{\chi'}_1$ have dimensions of mass. 
The parameters\\
$\lambda_{\Phi}, \lambda_{\eta}, \lambda_{S}, \lambda^{\chi}_{1,...4}, \tilde{\lambda}^{\chi}_{2,3}, \tilde{\lambda}^{\chi'}_5, \lambda^{(1)}_{\Phi \chi}, \lambda^{(2)}_{\Phi \chi}, \lambda^{(1)}_{\Phi \chi'}, \lambda^{\Phi \eta}_{1,2,3}, \lambda_{\Phi S}, \lambda^{(1)}_{\eta \chi}, \lambda^{(2)}_{\eta \chi}, \lambda^{(1)}_{\eta \chi'}, \lambda_{\eta S}, \lambda^{\chi \chi'}_{1,2,3}, \lambda_{\chi S}$ and  $\lambda_{\chi' S}$ are all dimensionless.
\section{Minimization of scalar potential}
\label{subsec:minimize}
The vacuum configuration leading to the breaking of discrete and electroweak symmetries are given by 
 \begin{eqnarray}
  \langle\Phi\rangle &=&
  {\left(\begin{array}{c}
  0 \\
  v_{\Phi}e^{i\theta}
 \end{array}\right)}~,\quad
 \langle\eta\rangle =0~,\quad
 \langle\chi_{1}\rangle= v_{\chi_{1}}e^{i\phi_{1}}~,\quad
 \langle\chi_{2}\rangle=v_{\chi_{2}}e^{i\phi_{2}}~,\nonumber \\
 &&\langle\chi_{3}\rangle=v_{\chi_{3}}e^{i\phi_{3}}~,\quad
  \langle\chi'_{1}\rangle=v_{\chi'_{1}}~,\quad
 \langle\chi'_{2}\rangle=v_{\chi'_{2}}~,\quad
 \langle\chi'_{3}\rangle=v_{\chi'_{3}}~,
  \label{neuvevs}
\end{eqnarray}
where $v_\Phi, v_{\chi_{1,2,3}}, v_{\chi'_{1,2,3}}$ are real and positive, and $\phi_{1,2,3}$ are physically meaningful phases. Since $\theta$ is not a physical 
observable, we set it to zero. Then there are ten minimization conditions for seven VEVs and three phases, using which we will find the vacuum configurations under the condition that the derivative of the potential $V$ given in Eq.(\ref{eq:pot}) with respect to each component of the scalar fields cease to exist for $\langle\eta\rangle=0$.
Starting  from the derivative of the potential with respect to the VEV of the SM Higgs filed we have
\begin{align}
v_{\Phi}^{2} &=
-\frac{1}{2 \lambda_{\Phi}}
\Big[
\mu_{\Phi}^{2}
+ \lambda_{\Phi\chi}^{(1)} \left( v_{\chi_i}^{2} + v_{\chi_j}^{2} + v_{\chi_k}^{2} \right) \notag\\
&\qquad
+ 2\, \lambda_{\Phi\chi}^{(2)} \left(
v_{\chi_i}^{2} \cos 2\phi_i
+ v_{\chi_j}^{2} \cos 2\phi_j
+ v_{\chi_k}^{2} \cos 2\phi_k
\right) \notag\\
&\qquad
- \lambda_{\Phi\chi'}^{(1)} \left(
v_{\chi'_i}^{2}
+ v_{\chi'_j}^{2}
+ v_{\chi'_k}^{2}
\right)
\Big]
\label{v-phi-i}
\end{align}
Similarly, from the components of the scalar field $\chi$, we get the following vacuum configuration,
\begin{equation}
\begin{aligned}
v_{\chi_i}^2 = &
-\frac{1}{
4 \Big[ \big(\lambda_{1}^{\chi} + \lambda_{2}^{\chi}\big) \cos 4\phi_i
       + \tilde{\lambda}_{2}^{\chi} \cos 2\phi_i \Big]
}
\Bigg\{
v_{\Phi}^2 \Big( \lambda_{\Phi\chi}^{(1)}
        + 2\, \lambda_{\Phi\chi}^{(2)} \cos 2\phi_i \Big) + 2\, \mu_{\chi}^2 \cos 2\phi_i
\\[4pt]
&\quad
+ m_{\chi}^2
+ \Big[
\big(v_{\chi'_i}^2 + v_{\chi'_j}^2 + v_{\chi'_k}^2 \big)\lambda_{1}^{\chi\chi'}
+ \big(v_{\chi'_j}^2 + v_{\chi'_k}^2 \big)\lambda_{2}^{\chi\chi'}
+ \big(v_{\chi'_i}^2 + v_{\chi'_k}^2 \big) \lambda_{3}^{\chi\chi'}
\Big]
\Bigg\} \neq 0~,
 \\[6pt]
&\langle\chi_{j}\rangle=\langle\chi_{k}\rangle=0~.
\end{aligned}
\label{v-chi-i}
\end{equation}
In the equations (\ref{v-phi-i}) and (\ref{v-chi-i}), $i,j,k=1,2,3~(i\neq j\neq k)$.
For the sake of simplification, in correlation with equations (\ref{v-phi-i}) and (\ref{v-chi-i}), we can adopt the following vacuum alignments for fields $\chi$ and $\chi'$:
\begin{eqnarray}
\langle \chi \rangle &=& v_{\chi_i} \, e^{i \phi_i} \, a_i ,
\quad
\text{with } a_1 = (1,0,0), \; a_2 = (0,1,0), \; a_3 = (0,0,1) \nonumber \\[6pt]
\langle \chi' \rangle &=& v_{\chi'} (1,1,1).
\label{eq:chip-vev}
\end{eqnarray}
Now according to the vacuum alignment shown in Eq.(\ref{eq:chip-vev}), the minimal condition with respect to $\phi_i$ can be obtained as follows
\begin{equation}
   -\frac{1}{4} \frac{\partial V}{\partial \phi_i}
= v_\chi^2 \left[
v_\Phi^2  \lambda_{\Phi\chi}^{(2)}  + \mu_\chi^2 +4 (\lambda_1^\chi + \lambda_2^\chi) v_\chi^2 \cos 2\phi_i
+ \tilde{\lambda}_2^\chi v_\chi^2
\right] \sin 2\phi_i = 0,
\end{equation}
and $\frac{\partial V}{\partial \phi_{j}}= \frac{\partial V}{\partial \phi_{k}}=0$ holds true by default with $i,j,k=1,2,3$ $(i\neq j\neq k)$. In addition, the minimization condition for VEV of field $\chi'$ is reduced and the corresponding vacuum configuration can be obtained from $\frac{\partial V}{\partial v_{\chi'}}=0$.

So considering the vacuum alignment $\langle\chi\rangle=v_{\chi}e^{i\phi}(1,0,0)$ where $v_{\chi}\equiv v_{\chi_{1}}$ and $\phi\equiv \phi_{1}$, and  $v_{\chi'}=v_{\chi'_1}=v_{\chi'_2} = v_{\chi'_3}$ the overall vacuum configurations can be written as follows
\begin{equation}
v_\Phi^2 =-~\frac{\mu_\Phi^2 + v_\chi^2 \left( \lambda_{\Phi\chi}^{(1)} + 2 \lambda_{\Phi\chi}^{(2)} \cos 2\phi  \right) + 3 \lambda_{\Phi \chi'}^{(1)} v_{\chi'}^2 }{2\lambda_\Phi}~,
\end{equation}
\begin{equation}
v_\chi^2 = -~\frac{
v_\Phi^2 \left( \lambda_{\Phi\chi}^{(1)} + 2 \lambda_{\Phi\chi}^{(2)} \cos 2\phi \right) \\
 + 2 \mu_\chi^2 \cos 2\phi + m_\chi^2 \\
+ v_{\chi'}^2 \left( 3 \lambda_1^{\chi\chi'} + 2 \lambda_2^{\chi\chi'} +  2\lambda_3^{\chi\chi'}  \right)}{4\left[ (\lambda_1^\chi + \lambda_2^\chi) \cos 4\phi + \tilde{\lambda}_2^\chi \cos 2\phi \right]}~,
\end{equation}

\begin{equation}
v_{\chi'}^2 =
-~\frac{3 \lambda_{\Phi \chi'}^{(1)} v_\Phi^2 + 3 m_{\chi'}^2 + v_\chi^2
\left( 3 \lambda_1^{\chi\chi'} + 2 \lambda_2^{\chi\chi'} +  2\lambda_3^{\chi\chi'}  \right)}
{ 24 \lambda_5^{\chi'}} - \frac{18 \xi_1^{\chi'}v_{\chi'}}{24 \lambda_5^{\chi'}}~,
\end{equation}
and the relevant scalar potential can be written as
\begin{align}
V_0 &= \mu_\Phi^2 v_{\Phi}^2
   + \lambda_\Phi v_{\Phi}^4
   + \lambda_{\Phi\chi}^{(1)} v_\Phi^2 v_\chi^2
   + 2 \lambda_{\Phi\chi}^{(2)} v_\Phi^2 v_\chi^2 \cos 2\phi \nonumber\\
&\quad + 3 m_{\chi'}^2 v_{\chi'}^2
   + 12 \lambda_5^{\chi'} v_{\chi'}^4
   + 12 \xi_1^{\chi'} v_{\chi'}^3
   + 3 \lambda_{\Phi \chi'}^{(1)} v_\Phi^2 v_{\chi'}^2 \nonumber\\
&\quad + 2 \mu^2_\chi v_\chi^2 \cos 2\phi
   + m_\chi^2 v_\chi^2
   + 2 \lambda_1^{\chi} v_\chi^4 \cos 4\phi
   + 2 \lambda_2^{\chi} v_\chi^4 \cos 4\phi \nonumber\\
&\quad + 2 \tilde{\lambda}_2^{\chi} v_\chi^4 \cos 2\phi
   + 3 \lambda_1^{\chi \chi'} v_\chi^2 v_{\chi'}^2
   + 2 \lambda_2^{\chi \chi'} v_\chi^2 v_{\chi'}^2
   + 2 \lambda_3^{\chi \chi'} v_\chi^2 v_{\chi'}^2 .
\end{align}
The minimization with respect to $\phi$, the vacuum configurations take the following forms

{\bf (i)} for $\phi=0, \pm\pi$
 \begin{align}
v_\chi^2 &=
- \frac{
v_\Phi^2 \left( \lambda_{\Phi\chi}^{(1)} + 2  \lambda_{\Phi\chi}^{(2)}  \right)
+ 2 \mu_\chi^2  + m_\chi^2
+ v_{\chi'}^2 \left( 3 \lambda_1^{\chi\chi'} + 2 \lambda_2^{\chi\chi'} +  2\lambda_3^{\chi\chi'} \right)
}
{4 \left[ (\lambda_1^\chi + \lambda_2^\chi) + \tilde{\lambda}_2^\chi  \right]}
\nonumber \\
v_\Phi^2 &=
- \frac{
\mu_\Phi^2 + v_\chi^2 \left( \lambda_{\Phi\chi}^{(1)} + 2  \lambda_{\Phi\chi}^{(2)} \right)
+ 3 \lambda_{\Phi \chi'}^{(1)} v_{\chi'}^2
}
{2 \lambda_\Phi}
\label{VP1}
\end{align}
{\bf (ii)} for $\phi=\pm\pi/2$
 \begin{align}
v_\chi^2 &=
- \frac{
v_\Phi^2 \left( \lambda_{\Phi\chi}^{(1)} - 2  \lambda_{\Phi\chi}^{(2)}  \right)
- 2 \mu_\chi^2  + m_\chi^2
+ v_{\chi'}^2 \left( 3 \lambda_1^{\chi\chi'} + 2 \lambda_2^{\chi\chi'} +  2\lambda_3^{\chi\chi'} \right)
}
{4 \left[ (\lambda_1^\chi + \lambda_2^\chi) - \tilde{\lambda}_2^\chi  \right]}
\nonumber \\
v_\Phi^2 &=
- \frac{
\mu_\Phi^2 + v_\chi^2 \left( \lambda_{\Phi\chi}^{(1)} -  \lambda_{\Phi\chi}^{(2)} \right)
+ 3 \lambda_{\Phi \chi'}^{(1)} v_{\chi'}^2
}
{2 \lambda_\Phi}
\label{VP1}
\end{align}
{\bf (iii)} for $\cos2\phi=-\frac{v^{2}_{\Phi} \lambda_{\Phi\chi}^{(2)}  +\mu^{2}_{\chi}+v^{2}_{\chi}\tilde{\lambda}^{\chi}_{2}}{4v^{2}_{\chi}(\lambda^{\chi}_{1}+\lambda^{\chi}_{2})}$

\begin{align}
v_\chi^2 &=
\frac{
2\left(\lambda^{\chi}_{1}+\lambda^{\chi}_{2}\right)\left(
m_\chi^2 + v_\Phi^2 \lambda_{\Phi\chi}^{(1)} + v_{\chi'}^2 \left( 3 \lambda_1^{\chi\chi'} + 2 \lambda_2^{\chi\chi'} +  2\lambda_3^{\chi\chi'} \right)
\right)
- \tilde{\lambda}^{\chi}_{2} \left(
v_\Phi^2  \lambda_{\Phi\chi}^{(2)}  + \mu_\chi^2
\right)
}
{
\tilde{\lambda}^{\chi2}_{2} + 8(\lambda^{\chi}_{1}+\lambda^{\chi}_{2})^{2}
}
\nonumber\\[10pt]
v_\Phi^2 &=
\frac{
2 v_\chi^2 \left(\lambda^{\chi}_{1}+\lambda^{\chi}_{2}\right) \left(
\mu_\Phi^2 + v_\chi^2 \lambda_{\Phi\chi}^{(1)} + 3 \lambda_{\Phi \chi'}^{(1)} v_{\chi'}^2
\right)
- \left( \lambda_{\Phi\chi}^{(2)}  \right)
\left( \mu_\chi^2 + v_\chi^2 \tilde{\lambda}^{\chi}_{2} \right)
}
{
\left( \lambda_{\Phi\chi}^{(2)}  \right)^2
- 4 \lambda_\Phi \left(\lambda^{\chi}_{1} + \lambda^{\chi}_{2}\right)
}
\end{align}
The first case (i) corresponds to non CP violation of the vacuum configuration whereas the cases (ii) and (iii) clearly show the breaking of CP symmetry.
\section{The scalar spectrum}
\label{higgs-mass}
In our intended model, there are two Higgs doublet and six Higgs singlets. After flavour and electroweak symmetry breaking, we will have,
\begin{eqnarray}
  \Phi &=&
  {\left(\begin{array}{c}
  0 \\
  v_\Phi+h
 \end{array}\right)}~,\qquad
 \eta =
  {\left(\begin{array}{c}
  \eta^{+} \\
  h_{1}+iA_{1}
 \end{array}\right)},\nonumber\\
 \chi_{1}&=& (v_{\chi}+\chi_{01})e^{i\phi}~,\chi_{2}=\chi_{02}~,\chi_{3}=\chi_{03}~,\nonumber\\
 \chi'_{1}&=& (v_{\chi'}+\chi'_{01})~,\chi'_{2}=(v_{\chi'}+\chi'_{02})~,\chi'_{3}=(v_{\chi'}+\chi'_{03}), S=s_{01},
  \label{Higgs}
 \end{eqnarray}
 with the SM VEV $v_\Phi=174$ GeV, $\eta^{+}\equiv(\eta^{-})^{\ast}$, $\varphi^{\pm}=0$, and $A_{0}=0$. We have tried to present the physical scalar boson masses where the standard Higgs $h$ is mixed with $\chi_{0i}$ and $\chi'_{0i}$, not with $h_1, A_1$. The neutral Higgs boson mass matrix in the basis of $(h, h_1,A_1,\chi_{01}, \chi_{02}, \chi_{03},\chi'_{01}, \chi'_{02}, \chi'_{03},s_{01} )$ is given by,
\begin{equation}
    M^2_{\rm{neutral}}=\begin{pmatrix}
m_h^2 & 0 & 0 & m_{h\chi_1}^2 & 0 & 0 & m_{h\chi'_1}^2 & m_{h\chi'_2}^2 & m_{h\chi'_3}^2 & 0 \\
0 & m_{h_1}^2 & 0 & 0 & 0 & 0 & 0 & 0 & 0 & 0 \\
0 & 0 & m_{A_1}^2 & 0 & 0 & 0 & 0 & 0 & 0 & 0 \\
m_{h\chi_1}^2 & 0 & 0 & m_{\chi_1}^2 & m_{\chi_1\chi_2}^2 & m_{\chi_1\chi_3}^2 & m_{\chi_1\chi'_1}^2 & m_{\chi_1\chi'_2}^2 & m_{\chi_1\chi'_3}^2 & 0 \\
0 & 0 & 0 & m_{\chi_1\chi_2}^2 & m_{\chi_2}^2 & m_{\chi_2\chi_3}^2 & m_{\chi_2\chi'_1}^2 & m_{\chi_2\chi'_2}^2 & 0 & 0 \\
0 & 0 & 0 & m_{\chi_1\chi_3}^2 & m_{\chi_2\chi_3}^2 & m_{\chi_3}^2 & m_{\chi_3\chi'_1}^2 & 0 & m_{\chi_3\chi'_3}^2 & 0 \\
m_{h\chi'_1}^2 & 0 & 0 & m_{\chi_1\chi'_1}^2 & m_{\chi_2\chi'_1}^2 & m_{\chi_3\chi'_1}^2 & m_{\chi'_1}^2 & m_{\chi'_1\chi'_2}^2 & m_{\chi'_1\chi'_3}^2 & 0 \\
m_{h\chi'_2}^2 & 0 & 0 & m_{\chi_1\chi'_2}^2 & m_{\chi_2\chi'_2}^2 & 0 & m_{\chi'_1\chi'_2}^2 & m_{\chi'_2}^2 & m_{\chi'_2\chi'_3}^2 & 0 \\
m_{h\chi'_3}^2 & 0 & 0 & m_{\chi_1\chi'_3}^2 & 0 & m_{\chi_3\chi'_3}^2 & m_{\chi'_1\chi'_3}^2 & m_{\chi'_2\chi'_3}^2 & m_{\chi'_3}^2 & 0 \\
0 & 0 & 0 & 0 & 0 & 0 & 0 & 0 & 0 & m_S^2
\end{pmatrix}.
\end{equation}
Here, the mass parameters are given as follows:
\begin{equation}
   m_h^2= 2 \left( 6 v^{2} \lambda_{\Phi} + v_{\chi}^{2} \left(\lambda_{\phi\chi}^{ (1)} + 2  \lambda_{\Phi\chi}^{ (2)} \cos 2\phi\right) + 3 v_{\chi'}^{2} \lambda_{\Phi\chi'}^{(1)} + \mu_\Phi^2 \right),
\end{equation}
\begin{equation}
   m_{h_1}^2 =2 \left( v_{\chi}^{2} \lambda_{\eta\chi}^ {(1)} + 2 v_{\chi}^{2} \lambda_{\eta\chi}^ {(2)} \cos 2\phi + 3 v_{\chi'}^{2} \lambda_{\eta\chi'}^{(1)} + v^{2}_{\Phi}\left( \lambda^{\Phi\eta}_ 1 + \lambda^{\Phi\eta}_ 2 + 2\lambda^{\Phi\eta}_ 3\right) + \mu_{\eta}^{2} \right) ,
\end{equation}
\begin{equation}
    m_{A_1}^2 = 2 \left( v_{\chi}^{2} \lambda_{\eta\chi}^ {(1)} + 2 v_{\chi}^{2} \lambda_{\eta\chi}^ {(2)} \cos 2\phi + 3 v_{\chi'}^{2} \lambda_{\eta\chi'}^{(1)} + v^{2}_\Phi \left(\lambda^{\Phi\eta}_ 1 + \lambda^{\Phi\eta}_ 2 - 2  \lambda^{\Phi\eta}_3\right) + \mu_{\eta}^{2} \right),
\end{equation}
\begin{align}
m_{\chi_1}^2 &=
2 \Big(
m_{\chi}^{2}
+ 2 \mu_{\chi}^{2} \cos 2\phi
+ 12 v_{\chi}^{2} \left[(\lambda_{1}^{\chi} + \lambda_{2}^{\chi}) \cos 4\phi
+ \tilde{\lambda}_{2}^{\chi} \cos 2\phi \right] \notag \\
& \qquad
+ v_{\chi'}^{2} \left( 3\lambda_{1}^{\chi\chi'} + 2\lambda_{2}^{\chi\chi'}
+ 2\lambda_{3}^{\chi\chi'} \right)
+ v_{\Phi}^{2} \left( \lambda_{\Phi\chi}^{(1)}
+ 2\lambda_{\Phi\chi}^{(2)} \cos 2\phi \right)
\Big)
\end{align}
\begin{align}
m_{\chi_2}^2 = 2 \Bigg(
& m_{\chi}^{2} + 2\,\mu_{\chi}^{2}
+ v_{\chi}^{2} \Big[
\left(4\lambda_{4}^{\chi} - \tilde{\lambda}_{2}^{\chi} + 4\tilde{\lambda}_{3}^{\chi}\right)\notag \\
& \quad+ \left(4\lambda_{1}^{\chi} - 2\lambda_{2}^{\chi} + 8\lambda_{3}^{\chi}
- 4\lambda_{4}^{\chi} + 4 \tilde{\lambda}_{3}^{\chi} - \tilde{\lambda}_{2}^{\chi}\right) \cos 2\phi
  + \sqrt{3}\,\tilde{\lambda}_{2}^{\chi} \sin 2\phi \Big]\notag \\
&+ v_{\chi'}^{2} \left( 3\lambda_{1}^{\chi\chi'} + 2\lambda_{2}^{\chi\chi'} + 2\lambda_{3}^{\chi\chi'} \right)
+ v_{\Phi}^{2} \left( 2\lambda_{\Phi\chi}^{(1)} + 2\lambda_{\Phi\chi}^{(2)} \right)
\Bigg),
\end{align}
\begin{align}
m_{\chi_3}^2 = 2 \Bigg(
& m_{\chi}^{2} + 2\,\mu_{\chi}^{2}
+ v_{\chi}^{2} \Big[
\left( 4\lambda_{4}^{\chi} + 4\tilde{\lambda}_{3}^{\chi} - \tilde{\lambda}_{2}^{\chi} \right) \notag \\
& \quad + \left( 4\lambda_{1}^{\chi} - 2\lambda_{2}^{\chi} + 8\lambda_{3}^{\chi}
+ 4\lambda_{4}^{\chi} + 4\tilde{\lambda}_{3}^{\chi} - \tilde{\lambda}_{2}^{\chi} \right) \cos 2\phi
- \sqrt{3}\,\tilde{\lambda}_{2}^{\chi} \sin 2\phi \Big] \notag \\
& + v_{\chi'}^{2} \left( 3\lambda_{1}^{\chi\chi'} + 2\lambda_{2}^{\chi\chi'} + 2\lambda_{3}^{\chi\chi'} \right)
+ v_{\Phi}^{2} \left( \lambda_{\Phi\chi}^{(1)} + 2\lambda_{\Phi\chi}^{(2)} \right)
\Bigg),
\end{align}
\begin{equation}
    m_{\chi_1'}^2 =
2 \left(
m_{\chi'}^{2}
+ v_{\chi}^{2} \left( \lambda_{1}^{\chi\chi'} + 2\lambda_{3}^{\chi\chi'} \right)
+ 8\, v_{\chi'}^{2} \lambda_{5}^{\chi'}
+ v_{\Phi}^{2} \lambda_{\Phi\chi'}^{(1)}
\right),
\end{equation}
\begin{equation}
    m_{\chi_2'}^2 =
2 \left(
m_{\chi'}^{2}
+ v_{\chi}^{2} \left( \lambda_{1}^{\chi\chi'} + \lambda_{2}^{\chi\chi'} \right)
+ 8\, v_{\chi'}^{2} \lambda_{5}^{\chi'}
+ v_{\Phi}^{2} \lambda_{\Phi\chi'}^{(1)}
\right),
\end{equation}
\begin{equation}
    m_{\chi_3'}^2 =
2 \left(
m_{\chi'}^{2}
+ v_{\chi}^{2} \left( \lambda_{1}^{\chi\chi'} + \lambda_{2}^{\chi\chi'} \right)
+ 8\, v_{\chi'}^{2} \lambda_{5}^{\chi'}
+ v_{\Phi}^{2} \lambda_{\Phi\chi'}^{(1)}
\right),
\end{equation}
\begin{equation}
    m_S^2 = 2\left( 3  v_{\chi'}^{2} \lambda_{\chi' S} + v_{\chi}^{2} \lambda_{\chi S}+ v_{\Phi}^{2} \lambda_{\Phi S} + \mu_{S}^{2}\right),
\end{equation}
\begin{equation}
    m_{h\chi_1}^2 = 4 v_{\Phi} v_{\chi} \left( \lambda_{\Phi\chi}^{(1)} + 2\lambda_{\Phi\chi}^{(2)}  \cos 2\phi\right),
\end{equation}
\begin{equation}
    m_{h\chi'_1}^2 =m_{h\chi'_2}^2=m_{h\chi'_3}^2= 4 v_{\Phi} v_{\chi'} \lambda_{\Phi\chi'}^{(1)},
\end{equation}
\begin{equation}
    m_{\chi_1\chi_2}^2 = 2 v_{\chi'}^2 \left( \left( \lambda_{2}^{\chi\chi'} - \lambda_{3}^{\chi\chi'}\right) \cos \phi - \sqrt{3} \lambda_{3}^{\chi\chi'}\sin \phi\right),
\end{equation}
\begin{equation}
    m_{\chi_1\chi_3}^2 = 2 v_{\chi'}^2 \left( \left( \lambda_{2}^{\chi\chi'} -\lambda_{3}^{\chi\chi'}\right) \cos \phi   +\sqrt{3} \lambda_{3}^{\chi\chi'}\sin \phi\right),
\end{equation}
\begin{equation}
    m_{\chi_1\chi'_1}^2=4 v_\chi v_{\chi'}\left( \lambda_{1}^{\chi\chi'}+ 2 \lambda_{3}^{\chi\chi'}\right),
\end{equation}
\begin{equation}
    m_{\chi_1\chi'_2}^2=m_{\chi_1\chi'_3}^2=4 v_\chi v_{\chi'}\left( \lambda_{1}^{\chi\chi'}+\lambda_{2}^{\chi\chi'}\right),
\end{equation}
\begin{equation}
    m_{\chi_2\chi_3}^2= 2 v_{\chi'}^2 \left(  \lambda_{2}^{\chi\chi'} - \lambda_{3}^{\chi\chi'}\right) +12 v_\chi \left(\xi^{\chi}_1 + \tilde{\xi}^{\chi}_1 \right) \cos \phi~,
\end{equation}
\begin{equation}
    m_{\chi_2\chi'_1}^2=m_{\chi_2\chi'_2}^2= 2 v_\chi v_{\chi'}\left(\left(\lambda_{2}^{\chi\chi'} -\lambda_{3}^{\chi\chi'}\right) \cos \phi-\sqrt{3} \lambda_{3}^{\chi\chi'}\sin \phi \right),
\end{equation}
\begin{equation}
    m_{\chi_3\chi'_1}^2=m_{\chi_3\chi'_3}^2= 2 v_\chi v_{\chi'}\left(\left(\lambda_{2}^{\chi\chi'} -\lambda_{3}^{\chi\chi'}\right) \cos \phi + \sqrt{3} \lambda_{3}^{\chi\chi'}\sin \phi \right),
\end{equation}
\begin{equation}
    m_{\chi'_1\chi'_2}^2 = m_{\chi'_1\chi'_3}^2 = m_{\chi'_2\chi'_3}^2 = 16 v_{\chi'}^2 \lambda^{\chi'}_5 + 12 v_{\chi'} \xi^{\chi'}_1.
\end{equation}
The mass parameter corresponding to the charged Higgs boson $\eta^{+}$ is given by,
\begin{equation}
    m_{\eta^+}^2 =
2 \left(
v_{\chi}^2 \left( \lambda_{\eta\chi}^{(1)} + 2\,\lambda_{\eta\chi}^{(2)} \cos 2\phi \right)
+ 3 v_{\chi'}^{2} \lambda_{\eta\chi'}^{(1)}
+ v_{\Phi}^{2} \lambda^{\Phi\eta}_{1}
+ \mu_{\eta}^{2}
\right).
\end{equation}
Using $m_{h_1}^2$, $m_{A_1}^2$, the expressions for $\bar{m}^{2}_{\eta}$ is given by,
\begin{align}
\bar{m}^{2}_{\eta} &=
2 v_{\chi}^2  \lambda_{\eta\chi}^{(1)}
+ 4 v_{\chi}^2 \lambda_{\eta\chi}^{(2)} \cos 2\phi
+ 6 v_{\chi'}^{2} \lambda_{\eta\chi'}^{(1)}
+ 2 v^{2}_{\Phi} \left( \lambda^{\Phi\eta}_1 + \lambda^{\Phi\eta}_2 \right)
+ 2\mu_{\eta}^{2} \\
&= m_{\eta^+}^2 + 2 v^{2}_{\Phi} \lambda^{\Phi\eta}_2.
\end{align}
\section{Light neutrino mass matrix}
\label{light-nu-mass}
The light neutrino mass matrix can be written as
\begin{equation}
 m_\nu = \begin{pmatrix}
          m_{11} & m_{12} & m_{13}\\
          m_{21} & m_{22} & m_{23}\\
          m_{31} & m_{32} & m_{33}
         \end{pmatrix}.
\end{equation}

Here
\begin{align*}
m_{11} &= m_0\, A\, y_1^2, &
m_{12} &= m_{21} = m_0\, B\, y_1 y_2, &
m_{13} &= m_{31} = m_0\, B\, y_1, \nonumber\\
m_{22} &= D\, y_2^2, &
m_{23} &= m_{32} = m_0\, G\, y_2, &
m_{33} &= m_0\, D.
\end{align*}
The values of $A, B, D, G ~\rm{and}~ m_0$ are defined in Eq.(\ref{entries}).
\begin{figure}[htbp!]
    \centering
    \subcaptionbox{\label{fig:m11}}{\includegraphics[width=0.3\linewidth,height=5.2cm]{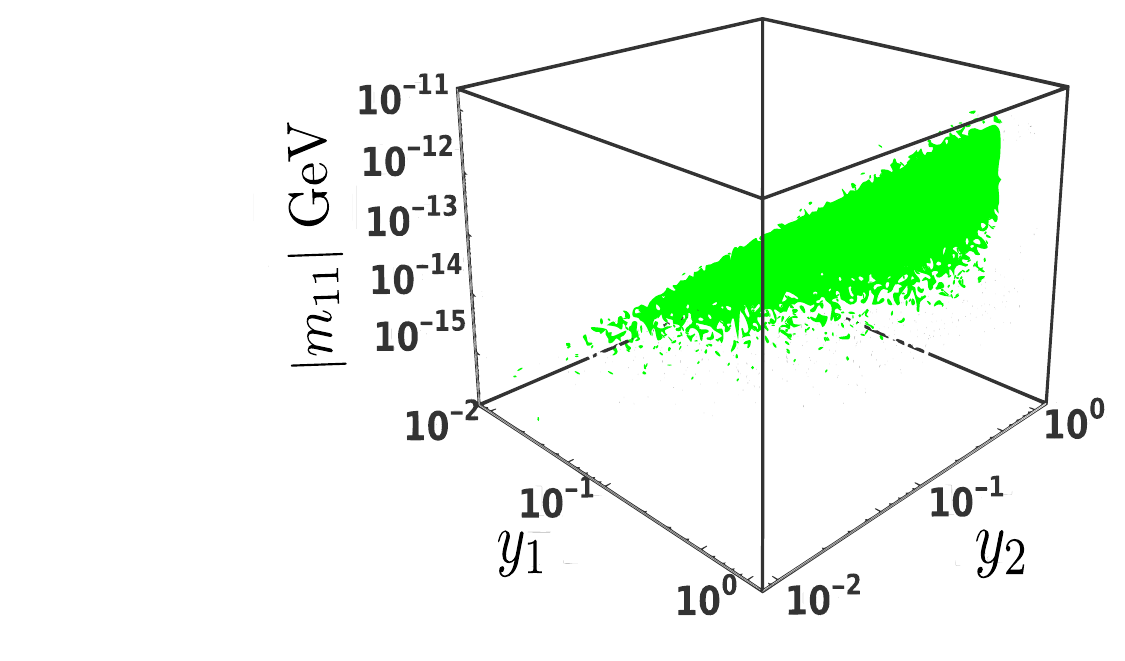}}
    \hspace{0.2cm}
    \subcaptionbox{\label{fig:m12}}{\includegraphics[width=0.3\linewidth,height=5.2cm]{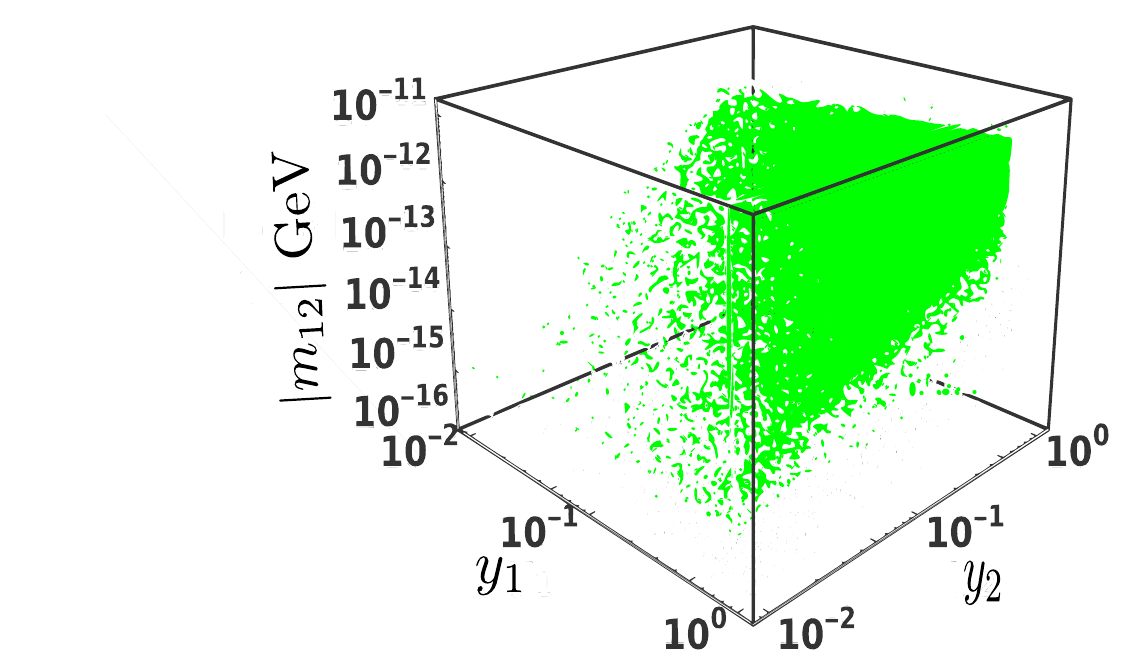}}
      \hspace{0.2cm}
    \subcaptionbox{\label{fig:m13}}
      {\includegraphics[width=0.3\linewidth,height=5.2cm]{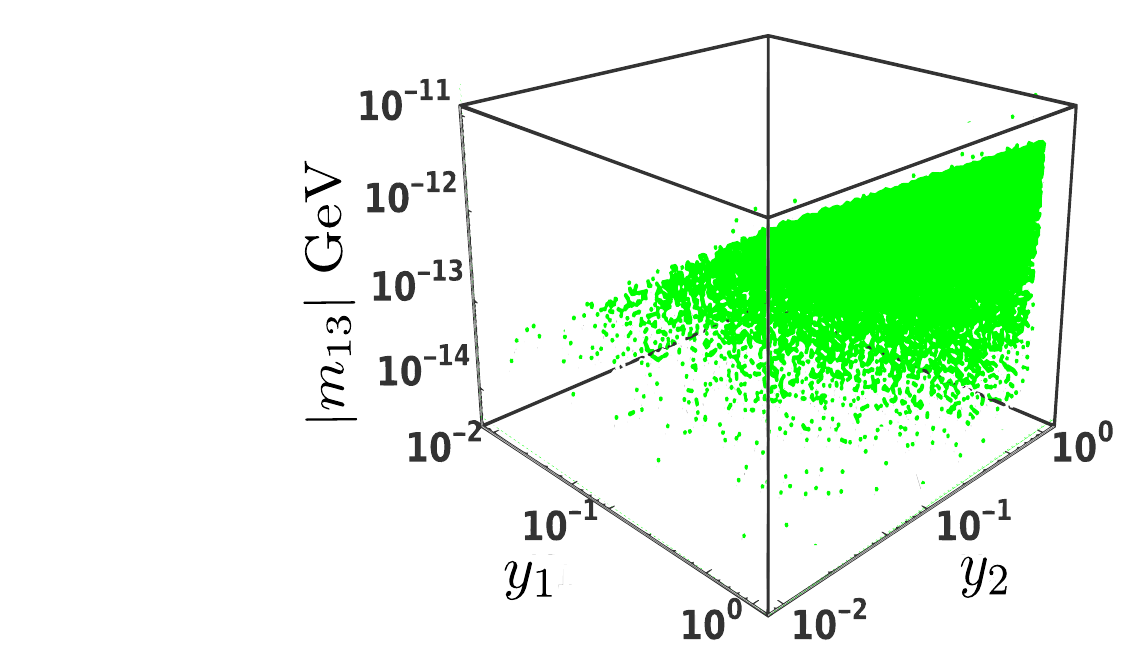}}\\
    \subcaptionbox{\label{fig:m22}}{\includegraphics[width=0.3\linewidth,height=5.2cm]{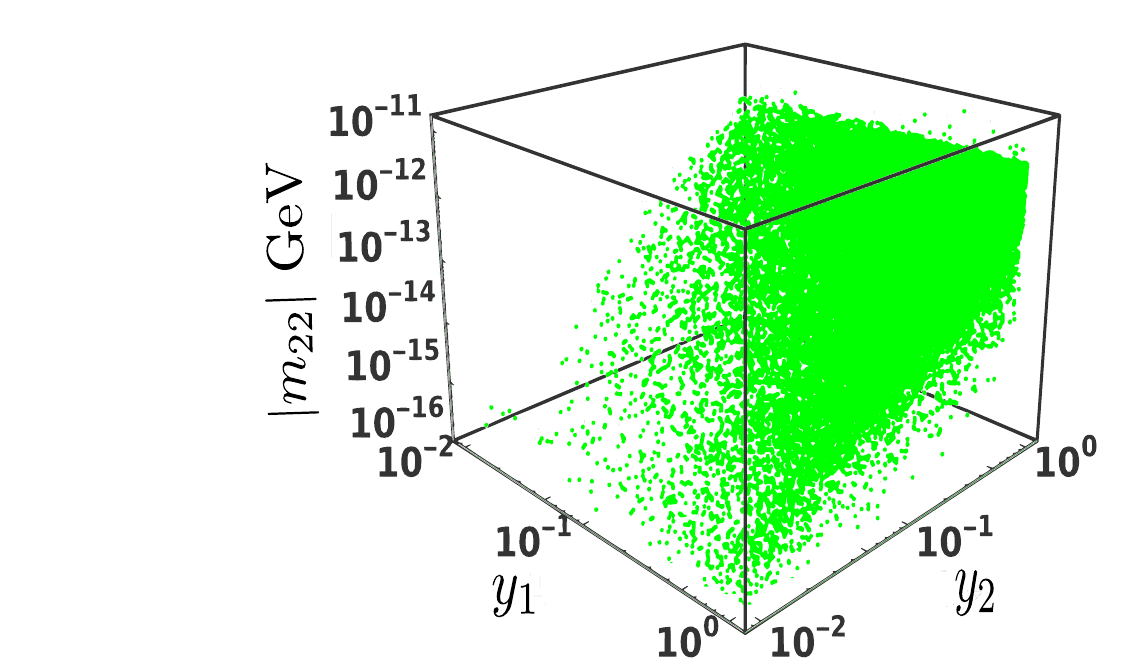}}
      \hspace{0.2cm}
    \subcaptionbox{\label{fig:m23}}
      {\includegraphics[width=0.3\linewidth,height=5.2cm]{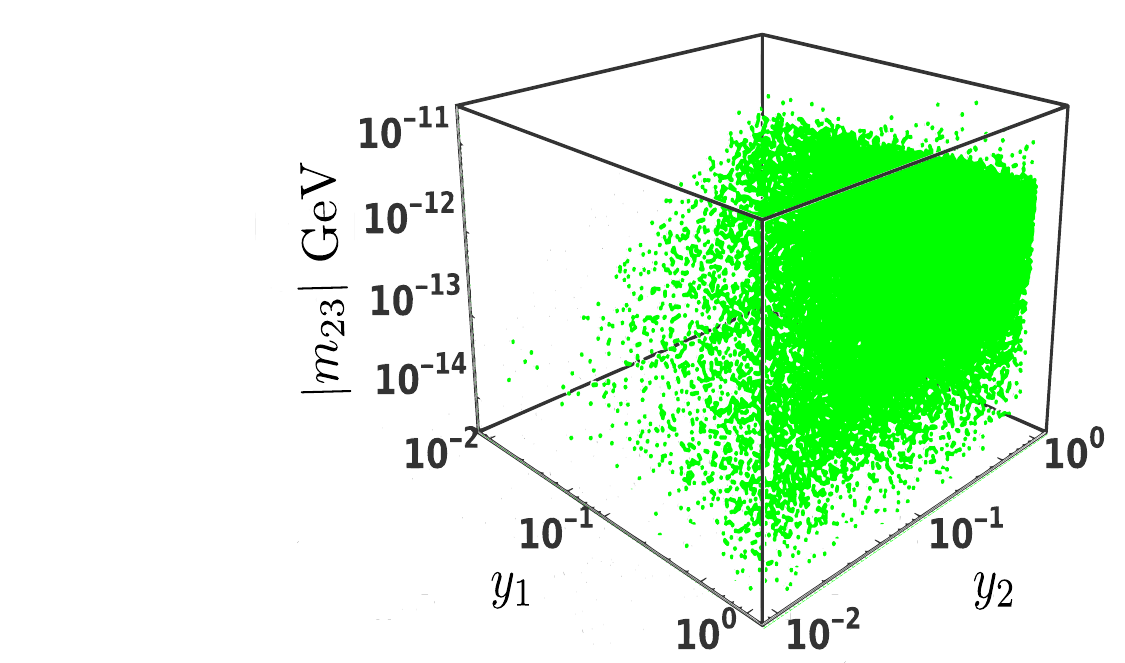}}
    \hspace{0.2cm}
    \subcaptionbox{\label{fig:m33}}{\includegraphics[width=0.3\linewidth,height=5.2cm]{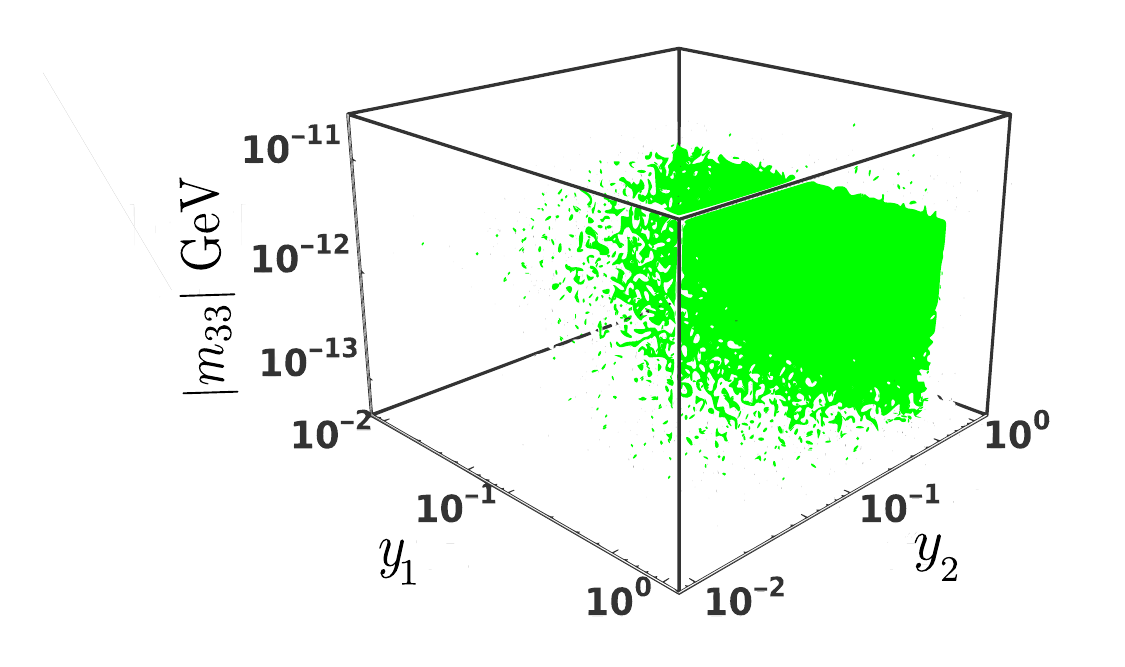}}
    \caption{The figure shows the range of each element of the light neutrino mass matrix using Eq.(\ref{radseesaw}). The expression for each element in terms of our model parameters is presented in section (\ref{sec:nu-matrix}). Subjected to the oscillation parameters for NH shown in the Table. (\ref{osci-parameter}) and using the operational range for our model parameter with associated restricting conditions (provided in section (\ref{result})), we are varying the matrix elements with respect to the Yukawa parameters $y_{1,2}$ and the upper limit of the matrix elements is found to be of the order $10^{-11}$ GeV. }
    \label{fig:light-nu-range}
\end{figure} 

\bibliographystyle{JHEP}

\end{document}